\documentclass{acmart}
\newcommand{\ourmodel}{\textit{ExpresSense}}
\newcommand{\ourmodelv}{\textit{ExpresSense}}

\usepackage{wrapfig}

\usepackage{amsmath}
\usepackage{amsfonts}
\usepackage{caption}
\usepackage{subfig}
\usepackage{color}

\usepackage{mathtools, cuted}
\usepackage{float}
\usepackage{array}

\usepackage{graphicx}
\graphicspath{{images/}{../images/}}
\usepackage{subfiles}
\usepackage{enumitem}
\newlist{steps}{enumerate}{1}
\setlist[steps, 1]{leftmargin=1.2cm, label = Step \arabic*:}

\usepackage{graphics}       
\definecolor{steelblue}{rgb}{0.27, 0.51, 0.71}

\newcommand{\revise}[1]{\textcolor{black}{#1}}

\renewcommand{\textrightarrow}{$\rightarrow$}

\usepackage{microtype}       

\usepackage{ccicons}          

\usepackage{longtable}
\usepackage{subfig}

\AtBeginDocument{%
  \providecommand\BibTeX{{%
    \normalfont B\kern-0.5em{\scshape i\kern-0.25em b}\kern-0.8em\TeX}}}

\begin{document}

\title[Exploring a Standalone Smartphone to Sense Engagement]{\textit{ExpresSense}: Exploring a Standalone Smartphone to Sense Engagement of Users from Facial Expressions Using Acoustic Sensing}

\author{Pragma Kar}
\email{pragyakar11@gmail.com}
\affiliation{%
  \institution{Jadavpur University}
  \country{India}
}
\author{Shyamvanshikumar Singh}
\affiliation{%
  \institution{Indian Institute of Technology Kharagpur}
  \city{Kharagpur}
  \country{India}
}
\email{shyamvanshikumar@gmail.com}

\author{Avijit Mandal}
\affiliation{%
  \institution{Indian Institute of Technology Kharagpur}
  \city{Kharagpur}
  \country{India}
}
\email{avijitmandal2001@iitkgp.ac.in}

\author{Samiran Chattopadhyay}
\affiliation{%
  \institution{Jadavpur University \&}
  \institution{TCG CREST}
    \country{India}}
\email{samirancju@gmail.com}

\author{Sandip Chakraborty}
\affiliation{%
  \institution{Indian Institute of Technology Kharagpur}
  \city{Kharagpur}
  \country{India}
}
\email{sandipchkraborty@gmail.com}

\begin{abstract}
Facial expressions have been considered a metric reflecting a person’s engagement with a task. While the evolution of expression detection methods is consequential, the foundation remains mostly on image processing techniques that suffer from occlusion, ambient light, and privacy concerns. In this paper, we propose \ourmodel, a lightweight application for standalone smartphones that relies on near-ultrasound acoustic signals for detecting users’ facial expressions. \ourmodel{} has been tested on different users in lab-scaled and large-scale studies for \revise{both posed as well as natural expressions}. By achieving a classification accuracy of $\approx 75\%$ over various basic expressions, we discuss the potential of a standalone smartphone to sense expressions through acoustic sensing. 
\end{abstract}



\maketitle

\section{Introduction}
\revise{Imagine an interactive smartphone application that can ``sense'' its user's mood when browsing social media profiles, watching a movie, or typing a long message on it. Considering the global prevalence of depressive and anxiety disorders, particularly among the young population~\cite{hawes2022increases}, such an application can pervasively help and alert individuals early. It can even recommend remedies, such as suggesting music, a funny reel, or a comedy movie that helps change the mood. Facial expressions are one of the prominent ways to correctly infer an individual's mood~\cite{10.1145/3316782.3321543} and thus can fulfill the above vision if the smartphone can monitor the temporal changes of its user's facial expressions. Interestingly, there have been decades of research on inferring facial expressions from video or image-based data~\cite{lisetti2000automatic,miao2019deep,zhao2021robust,ben2021video}; however, these works are not suitable to fulfill the above vision of developing a pervasive smartphone application because of the following reasons.} \textbf{Firstly}, image and video processing is computationally heavy and consumes a significant amount of system resources and energy. Running a computationally heavy model on a hand-held device like a smartphone is not always feasible as it will affect the device's performance. \textbf{Secondly}, in the absence of proper lighting conditions, camera-based techniques result in missed detection. In a scenario where a person is watching a movie while sitting in a dark room, camera-based expressions detection will fail due to the lack of ambient light. \textbf{Thirdly}, the most important drawback of image and video-based systems is the privacy concern. In systems that aim to monitor all user activities continuously and operate as a ``watchdog'', the users often feel uncomfortable. Moreover, continuous camera usage also depletes the battery life at an unusually faster rate.

To achieve the above vision, in this paper, we explore acoustic sensing over a commercial off-the-shelf (COTS) smartphone to identify four basic facial expressions of the user when they browse through a smartphone app (say, watching a movie on Netflix). Due to smartphones' relatively lower processing capabilities, lightweight solutions need to be developed for expression detection, and acoustic processing evokes less power consumption than image processing techniques. Since the trade-off between accuracy and system resource consumption should be optimal while designing expression detection models for smartphones, unlike that for desktops where accuracy is of prime importance, acoustic sensing becomes a suitable approach for smartphones. \revise{Summarily, developing}
\begin{wrapfigure}{l}{0.4\textwidth}
    \centering
    \includegraphics[width=0.9\linewidth]{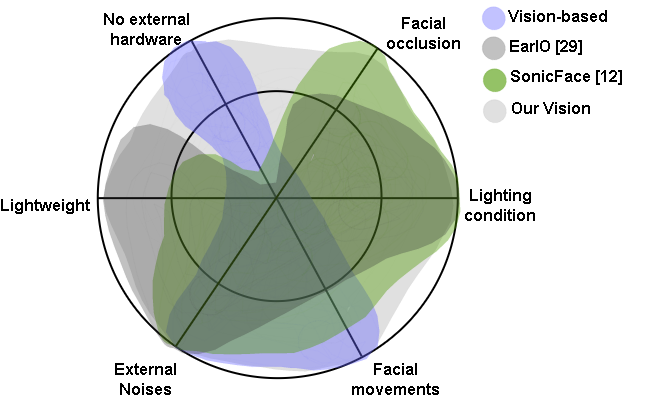}
    \caption{Our vision in contrast to the existing literature}
    \Description{A figure showing the six requirements of facial expression detection systems: resistant to facial occlusion, different lighting condition, facial movement, external noise, lightweight, and no external hardware. The figure contrasts the contribution of vision-based techniques, EarIO, and SonicFace in addressing these requirements and compares our vision with these approaches.}

    \label{fig:vision}
\end{wrapfigure}
\revise{a pervasive smartphone system for facial expression recognition has the following requirements -- (1) lightweight technology with in-device computing, (2) no use of external hardware, (3) performance-invariant under different lighting conditions, (4) correct detection of different facial movements, (5) should not get impacted from occlusions, like glasses or face masks, and (6) performance-invariant from external noises, like motion or interference from other signals. Interestingly, there have been a few recent works~\cite{gao2021sonicface,li2022eario} that explore acoustic sensing for tracking facial movements. Vision-based approaches~\cite{miao2019deep} can capture a maximum number of facial expressions while capturing the video/image with the embedded camera of the device. However, both SonicFace~\cite{gao2021sonicface} and EarIO~\cite{li2022eario}, although work based on acoustic sensing principle, they need external hardware supports (while the former needs a microphone array, the latter works on earphones). Further, SonicFace is computationally heavy and typically works on a desktop computer, while EarIO needs specific hair arrangements (ponytail hairstyle). Further, EarIO captures only some specific facial movements (like eye, mouth open/close, and their combinations) and not the expressions directly. Figure \ref{fig:vision} shows how our vision bridges these gaps and addresses the basic requirements. }

\subsection{How Do We Utilize Acoustic Sensing?}
In general, visible facial expressions that last between $0.5$--$4$ seconds are broadly grouped under macro expressions. Apart from these, more subtle and spontaneous micro-expressions last for less than half a second~\cite{zong2019cross}. Irrespective of the type of facial expression, each expression results from a set of \textit{Action Units} (AU) (facial muscle movements). Interestingly, facial muscle movements are the building blocks of expressions and can be categorized under \textit{ocular} (around the eye region), \textit{nasal} (around the nose region), and \textit{oral} (around the mouth area) groups~\cite{songfacelistener}. For example, a particular expression, say ``\textit{Happiness}'', is a combination of facial muscle movements, mainly around the oral region and subtly around the ocular and nasal area. \revise{The core idea of this paper is to detect such AUs through lightweight acoustic sensing over a COTS smartphone to capture a subject's facial expression. In contrast to the complex signal processing techniques over microphone array as used in SonicFace~\cite{gao2021sonicface} or deep learning model used in EarIO~\cite{li2022eario}, we rely on simple signal processing techniques and light machine learning models that can effectively be implemented on a smartphone app, while using only the embedded smartphone hardware.}

\subsection{Our Contributions}
In the essence of the above discussion, we propose \ourmodel, a lightweight smartphone application that utilizes near-ultrasound acoustic signals to detect four basic expressions~\cite{Rachael,almeida2019altriras} of a user: \textit{Happiness}, \textit{Anger}, \textit{Surprise}, and \textit{Sadness} (or \textit{Neutral}\footnote{Existing literature argue that for both \textit{Sadness} and \textit{Neutral}, there is no visible sign of the facial gestures, and they appear to be very similar; therefore, these two expressions are used interchangeably in the literature~\cite{lee2008neutral,wang2018sad,ito2019emotional}.}). \ourmodel{} transmits chirps between the range of $16$-$19$ kHz, using the inbuilt speaker of a commercial smartphone. The reflected chirps are recorded through the single microphone of the smartphone, filtered, and processed to derive the amplitude and phase of the reflected signal for different frequency bins. The frequency bins that contribute the most to predicting the expressions are selected. This process of bin pruning is followed by utilizing the phase and amplitude of the echo from the desired frequency bins and using them to predict the corresponding expression of the subject by a pre-trained ensemble of classifiers. 

The significant challenges in developing \ourmodel{} arise from three primary aspects: design, development, and data. From the perspective of designing the system, the underlying characteristics of facial expressions, acoustic signals, and their intricate correlation needs to be considered. Assessing whether the facial expression has an identifiable effect on the signal reflection concerning the signal features poses a challenge in designing the proposed system. Next, a major challenge is to develop a system that overcomes the limitations of commodity smartphones. We address this issue by developing a model that can thrive on any commercial smartphone's very minimal and basic capability. The main contributions of this work are as follows.
\begin{enumerate}
    \item Development of a lightweight and camera-free smartphone application that uses acoustic signals for facial expression detection. The system works on a standalone smartphone and requires no additional hardware.
    \item Experimental analysis of \ourmodel{} and its application, demonstrating its significant performance under a realistic environment \revise{for both posed and natural expressions}.
\end{enumerate}  
We conduct a thorough lab-scaled study with $12$ participants to evaluate the performance of \ourmodel\footnote{We have taken the institute’s ethical committee’s approval to perform all the human studies reported in this paper.} -- \revise{both as a standalone platform under a controlled environment as well as an embedded application in the wild}. The experimental results reveal that \ourmodel{} can work with an average accuracy of about $\approx75\%$ as a user-dependent model and performs significantly well under different conditions like angular variance, ambient sound, motion, and so on. \revise{Further, to evaluate \ourmodel{} under a realistic setup with natural expressions, we develop a smartphone app that can monitor and measure user engagement while watching a streaming video. The application matches the overall facial expressions of the subject and the video genre to provide a temporal variation in the engagement score, which we compare with the ground truth captured from questionnaires and self-assessment. From a thorough study with the $12$ participants under an in-the-wild setup, we observe that the app performs with an average F1-score of $0.84$.} Further, we performed a large-scale study with $72$ participants to test the usability of \ourmodel{} with the help of this video streaming app that monitors the users' engagement and shares a summary report with them at the end of the streaming. The study revealed a high usability score of $85$, indicating the system's effectiveness.

\label{intro}
\section{Related Work}
\label{RW}
In this section, we discuss how acoustic sensing has been used in various applications and how they have inspired the idea of \ourmodel. Acoustic signals have been used extensively in coarse-grained motion and location tracking of  objects~\cite{10.1145/3384419.3430780,garg2021owlet,cai2021we}, as well as tracking subtle movements like respiratory patterns of individuals~\cite{10.1145/3478123,wan2021resptracker}. In EchoSpot~\cite{lian2021echospot}, target localization is performed using Frequency Modulated Continuous Wave (FMCW) signals that get reflected in the microphone from different paths. The system can find an individual's location in an indoor scenario by processing the received echo. Apart from chirps, ultrasonic tones~\cite{lian2021fall} has also been used in course-grained motion detection, e.g., sudden falls, using techniques like Doppler Shift~\cite{sen2011roadsoundsense}. Fine-grained motion tracking facilitates several healthcare applications, such as monitoring an individual's chest wall~\cite{10.1145/3372224.3419209}. In~\cite{10.1145/3463521} Liu \textit{et al.} proposed a system called BlinkListener that operates on the reflection acoustic chirps to detect blinking instances of an individual. The authors have analyzed the correlation between blink-induced motion and the phase and amplitude of the signal, as captured through a smartphone's microphone. The system can find its application in the detection of early symptoms of eye diseases or traffic safety. Smartphone-based acoustic systems have also enhanced communication clarity by inducing acoustic signal-based lip movement tracking~\cite{zhang2021sensing}. Jin \textit{et al.} have proposed a system~\cite{10.1145/3494992} that uses a portable speaker and a smartphone to transmit and receive sound signals for detecting traffic conditions and ensuring the safety of cyclists. 

Acoustic sensing has also been used in weather monitoring. Cai \textit{et al.} \cite{cai2020acute} proposed a novel technique that explores the correlation between air temperature and the speed of sound. The system converts a commodity smartphone into a standard thermometer by estimating the time between the chirp generated from the phone's speaker and the echo received by dual microphones. Ultrasonic sound signals can also assist in various security purposes. SilentSign \cite{chen2020silentsign} assists in identifying authentic signatures by emitting sound signals from a smartphone near a pen. The reflected echo is then processed to estimate the pen's distance at consecutive instances, thus verifying a signature. In terms of authentication, Isobe \textit{et al.} proposed a system using smart glasses~\cite{10.1145/3460421.3480425} that uses sound, transmitted through nose pads, to authenticate them. Sound,  a rich source of information regarding daily life activities, has also been used in activity recognition by amplifying the functionality of voice assistants~\cite{10.1145/3448090}.  

In terms of the devices used, it is evident from the above discussion that smartphones have been used as a standard medium for transmitting and receiving acoustic signals, as they contain onboard microphones and speakers and are globally available. \revise{However, the existing models for smartphone-based acoustic sensing are limited to detecting significant movements of specific body parts, such as eyes or lips. In contrast, detecting facial expressions involves holistic sensing of various facial muscles, which is challenging and cannot be accomplished with the existing mechanisms as they work for significant movement detection from close proximity (such as $<30$ cm for BlinkListener~\cite{10.1145/3463521}) only.} However, complex movements of objects, including the detection of facial expressions, have been carried out using microphone arrays. For instance, commercial microphone arrays have been used to localize and differentiate between multiple acoustic sources~\cite{wang2020symphony,gerstoft2021audio} and facial expression detection~\cite{gao2021sonicface}. Apart from the microphone arrays, earphones have also been used for tracking movement~\cite{cao2020earphone}, \revise{ and monitoring facial muscle movements~\cite{10.1145/3534621} } pertaining to the flexible positioning of the speakers and microphones. Earphones have found their usage in sign language detection~\cite{jin2021sonicasl} by acting as an on-body acoustic sensor. Similarly, hand gestures have been classified using acoustic beamforming techniques~\cite{iravantchi2019beamband}, facilitated by dedicated wristbands. Smart glasses, coupled with microphones and speakers, have also been used for acoustic sensing. In~\cite{10.1145/3448105}, smart glasses are used to capture facial muscle movements and classify facial actions like cheek and brow raise, winks, etc.

\revise{From the above discussions, it is evident that although it is explored widely for different applications, acoustic sensing is limited to utilizing sophisticated hardware like a microphone array or an additional earphone to sense fine-grained movements. On the contrary, COTS smartphones contain a single microphone and thus pose several additional challenges when used as a medium for acoustic sensing. In the next section, we explore the opportunities and challenges of using a single COTS smartphone for facial expression detection.}  

\section{ExpresSense Design: Opportunities and Challenges}
We start with a pilot experiment to understand how the acoustic signal generated from a COTS smartphone gets impacted by the movement of facial muscles. By showing how near-ultrasound signals can detect such movement, we then discuss the challenges associated with a COTS smartphone in designing a system like \ourmodel. 

\subsection{Pilot Study}\label{chirpchar}
\ourmodel{} considers \textit{Frequency Modulated Continuous Wave} (FMCW) or chirps that have linearly increasing frequency in time. Based on the characteristic of transmitted and reflected signals, we can say that a reflected signal is only a time-delayed variation of the transmitted chirp. In~\cite{10.1145/3463521}, the authors explain how a change in the reflective surface, along with subtle movements causes a shift in both phase and amplitude of the signal. This is caused by both the nature of the reflective surface and the length of the path the transmitted signal travels before hitting the reflective surface. If the reflective surface is skin, some signals will be absorbed, causing higher attenuation. On the other hand, more reflective surfaces, like teeth, eyeballs, etc., will result in lower attenuation. This affects the amplitude of the signal. Therefore, the amplitude value of the signal when a person is laughing (teeth are visible)  or surprised (eyes enlarged) will be significantly different from when they are sad (mouth closed, eyes normal). The signal path is affected by actions like an eye blink when the eyelid comes before the eyeballs and reflects the incoming signal. This causes a change in the signal phase. In terms of facial expressions, the phase of a signal will be significantly different when a person is surprised (mouth opened) or laughing (lips separated) from a scenario when the person is sad (lips cover the teeth). 

\revise{Thus, in contrast to the existing approaches~\cite{gao2021sonicface,10.1145/3384419.3430780,garg2021owlet,cai2021we,lian2021echospot} that use complex frequency-domain signal processing techniques}, we consider the \revise{time-domain} amplitude and phase of a signal by \revise{intelligently and adaptively} choosing a frequency bin that can capture the facial expression information. For choosing the frequency bin, we rely upon the discussed characteristic of amplitude and phase variation. If there are moving objects in the environment, like ceiling fans, moving curtains or passing vehicles, the changes in amplitude of the signals from these mediums will be less due to the static reflective surface. Based on this understanding, we select the frequency bin with the largest variance in the phase. We next analyze how the amplitude and the phase of a signal varies due to the movement of facial muscles. 
\begin{figure}[!t]
\begin{minipage}[b]{.5\textwidth}
\includegraphics[trim={3.5cm 1cm 3.5cm 1.5cm},clip,width=\textwidth]{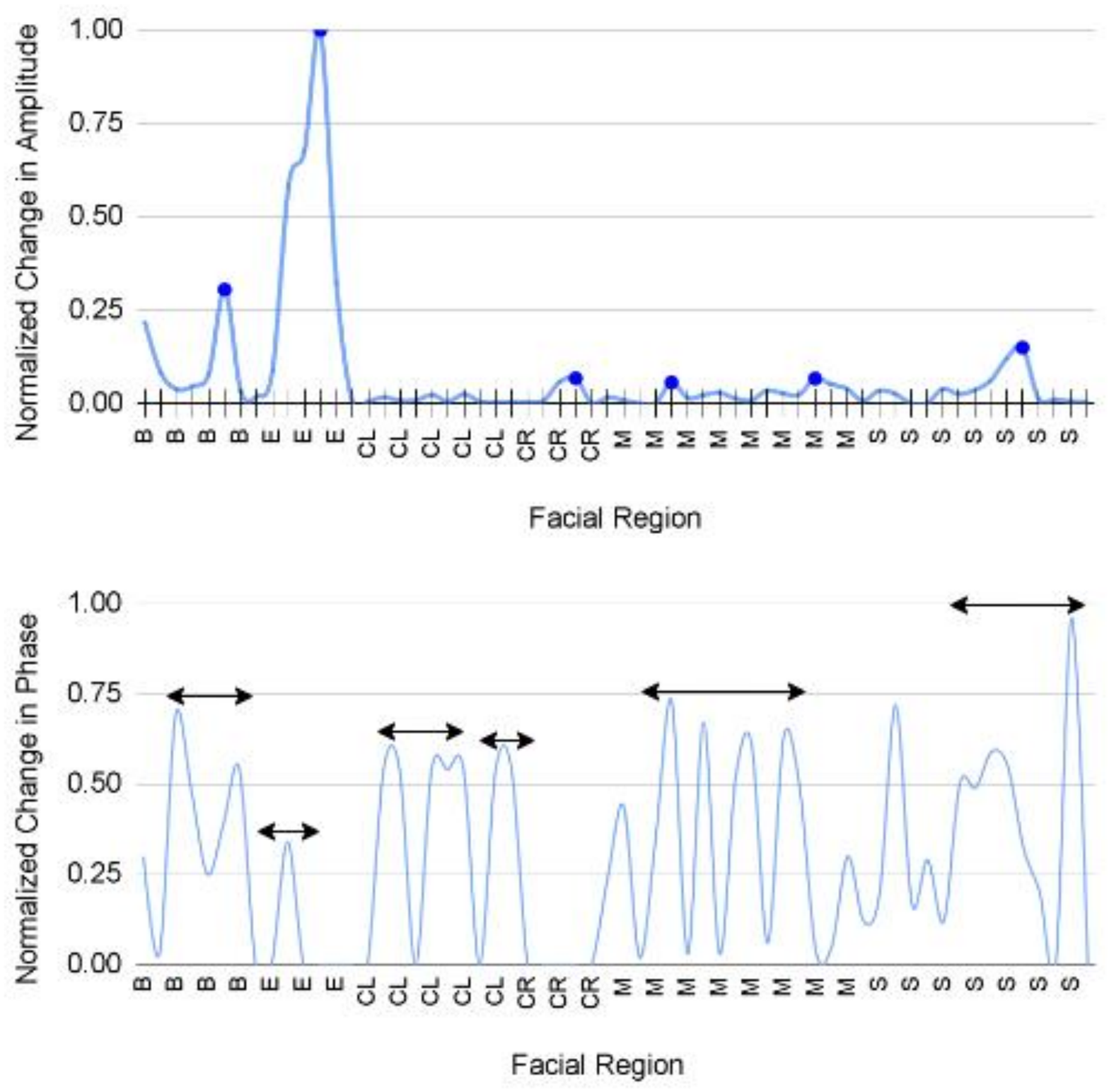}
\caption{Facial Muscles and Acoustic Feature Variation}
\Description{The figure shows that amplitude and phase of reflected sound signals change when facial action units like brows, cheeks, eyes, and moth regions are moved. }
\label{obs}
 \end{minipage}
  \hfill
  \begin{minipage}[b]{.45\textwidth}
\includegraphics[width=\textwidth]{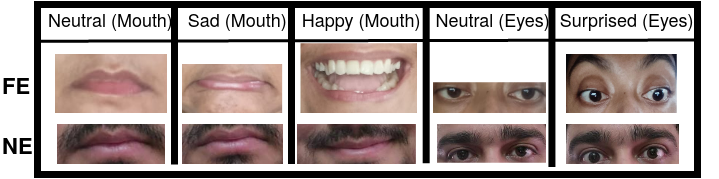}
\caption{Movement of facial muscles due to Forced Expressions (FE) and Natural Expressions (NE)}
\Description{The figure shows a  series of facial action units like moth and eyes for forced expressions and natural expressions. }
\label{posn}
  \end{minipage}
\end{figure}

\subsubsection{Methodology}
To test whether \revise{smartphone-generated} acoustic signals can differentiate between relaxation and contractions of facial muscles in different regions, we performed a pilot study with two subjects. We asked the subjects to perform the following sequence of facial actions while relaxing the facial muscles in between every two actions: \textit{Raise eyebrows/frown} (B), \textit{blink} (E), \textit{raise left cheek} (CL), \textit{raise right cheek} (CR), \textit{move mouth region (left and right)} (M), and \textit{smile} (S).  During the experiment, we placed a smartphone in front of their faces at a distance of about $30$ cm and an elevation and angle of zero degrees. We continuously played FMCW signals between $16$-$19$ kHz, through the speaker. The reflected signals are recorded through the microphone. After a series of processing (explained in Section~\ref{det}), we derive the phase and amplitude of the reflected chirps and select the frequency bin with the most significant variation, as mentioned earlier. We then plot the absolute difference in consecutive amplitudes and consecutive phases of these signals with respect to the time.

\subsubsection{Observations}\label{pilorobs}
The pilot study led to the following observation: (1) Facial AUs involving a significant muscle movement induce a change in either amplitude, phase, or both. (2) AUs that involve lower muscle movement and no change in the reflective surface do not affect the signal features. For example, the movement in the nasal region failed to change the amplitude and the phase in the selected face-bin. To further compare the degree of effect individual AU has on the phase and the amplitude, we plot Figure~\ref{obs}. We prune out the zones with no movement (face is relaxed) and nasal movement as they cause no variation in features. Then, we plot these variations for phase and amplitude individually, along with the corresponding facial AU that caused it. Figure~\ref{obs} shows that each AU has resulted in a peak in the plot. However, it can be seen that the blink has caused the most prominent peak, whereas the AUs around the oral region generated lower peaks. This observation infers that the amplitude and the phase of an FMCW signal reflection can be used to predict the facial expressions of a subject due to their varying characteristics and correlation with the underlying expressions.

\revise{Notably, we observe the above variations in the amplitude and phase values of the reflected signal for facial expressions that are posed.} In most popular image data sets, facial expressions are posed and quite different from more subtle expressions in real life. In Figure~\ref{posn}, the difference between posed or forced expressions and natural expressions is shown in terms of displacement around ocular and oral regions. It can be seen that forced expressions are much more animated, thus causing much more movement of the AUs. However, normal expressions that can be seen in daily life are more abstruse in terms of AUs. Therefore, in \ourmodel, we aim to estimate how smartphones can distinguish between these challenging natural expressions, by exploring the features (amplitude and phase) of the reflected signal and \revise{then learning the subtle variations in the signal properties through light-weight machine learning approaches}.

\subsection{Challenges and Design Ideas}\label{lim}
\revise{Although observing the time-domain signal properties for judicially selected frequency bins provides us an opportunity to develop a lightweight model for detecting facial expressions over smartphones, we still face the following challenges.}\\

\noindent\textbf{Supported Frequency Range:} \revise{Majority of the COTS smartphones work over the audible frequency range ($<20$ kHz). Moreover, signals above $19$ kHz are very noisy and unsuitable for the purpose of sensing. Consequently, in \ourmodel{}, we use a near-ultrasonic range of $16$-$19$ kHz to utilize the maximum possible bandwidth of $3$ kHz and ensure the least overlap with the audible range. Notably, even though the audible range varies between $20$ Hz to $20$ kHz, most of the audible sounds lie below $16$ kHz, thus ensuring minimal interference with \ourmodel{}.}\\

\noindent\textbf{Single Microphone:} \revise{The popular acoustic sensing approaches~\cite{gao2021sonicface,10.1145/3384419.3430780,garg2021owlet,cai2021we,lian2021echospot} use a microphone array to estimate the \textit{Direction of Arrival} (DoA) of the signal, thus eliminating the unwanted signal components. In contrast, most COTS smartphones use a single microphone, thus limiting the number of signal properties we can utilize. Further, approaches like SonicFace~\cite{gao2021sonicface} uses signal fusion technique to precisely track even minor object movements, which is not possible with a single microphone due to the lack of sophisticated interference cancellation techniques. Therefore, \ourmodel{} solely relies on lightweight, intelligent signal processing and machine learning methods to judicially select the frequency bins, which can eliminate signal components reflected from the background objects.}\\

\noindent\textbf{Lightweight:} \revise{The developed system should be lightweight to be deployed directly on the smartphone. Existing studies use complex frequency-domain signal processing techniques or computationally heavy deep learning models. Such methods will consume significant computational resources (like RAM) and thus can slow down other running services on the smartphone. So, we have to rely on lightweight techniques that need minimal computation.}

\section{The Overview of \ourmodel{}}

\begin{figure}[!t]
 \begin{minipage}[b]{0.45\textwidth}
\includegraphics[width=\textwidth]{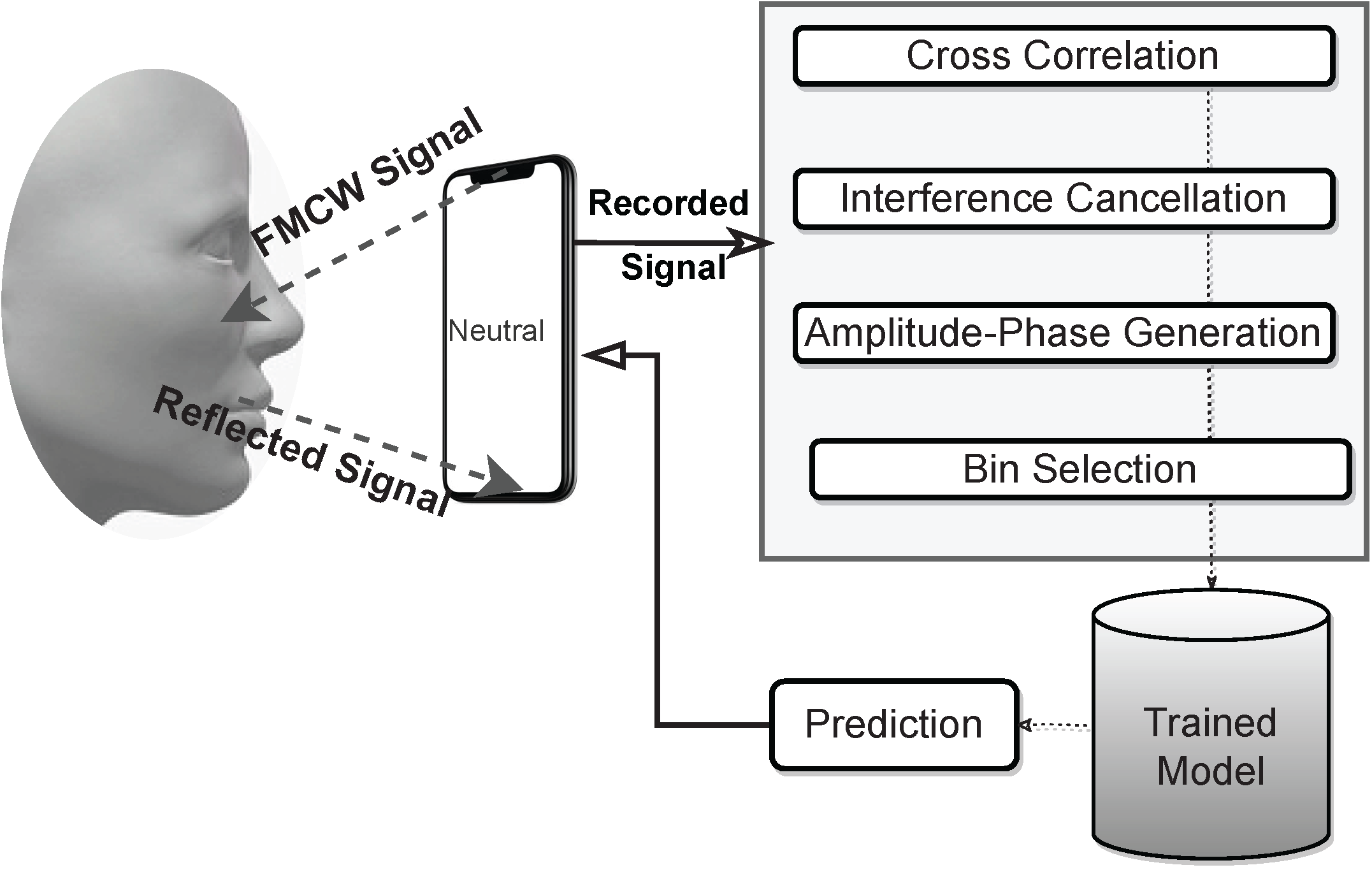}
\caption{The overview of \ourmodel}
\Description{The diagram shows the overview of \ourmodel. The application contains the following sub-modules: Cross-Correlation, Interference Cancellation, Amplitude-Phase Generation, Bin Selection, and Prediction by a trained model.  }
\label{overv}
\end{minipage}
  \hfill
  \begin{minipage}[b]{0.45\textwidth}
\includegraphics[width=.8\textwidth]{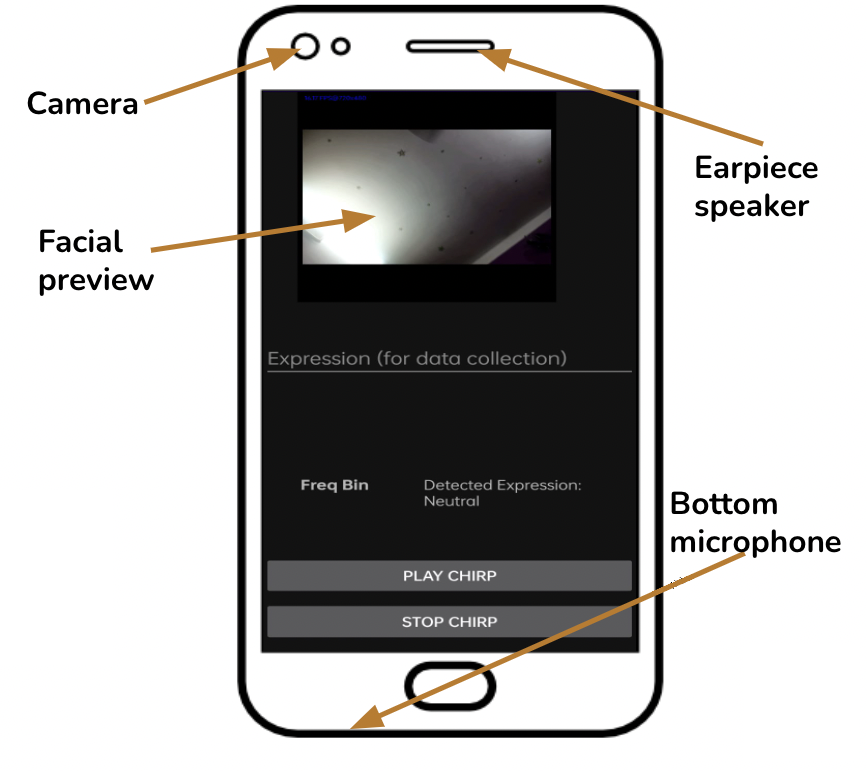}
\caption{ The interface of \ourmodel for data collection.}
\Description{The figure shows the interface of \ourmodel{} that contains a camera-preview, text box for manual label entry, play chirp, and stop chirp buttons. The figure also shows the location of the device's camera, speaker, and microphone.  }
\label{ver}
  \end{minipage}
\end{figure}

Figure~\ref{overv} shows the overview of \ourmodel{} in terms of chirp transmission, reception, and in-device processing. We start with discussing an overview of the proposed architecture, followed by a discussion on the smartphone applications developed for data collection and prediction, adhering to the proposed architecture.

\subsection{Proposed Architecture}
In \ourmodel, a standalone commodity smartphone is used for camera-less expression detection using near-ultrasound signals. The overall idea of \ourmodel{} is to generate near-inaudible chirp signals ranging from $16$-$19$ kHz utilizing the smartphone's speaker. The smartphone's microphone is used to capture the echo of the signals reflected from the facial regions of the subject, along with other objects in the subject's vicinity. The recorded echo is then processed to prune out the static multi-path interference. The amplitude and phase from different frequency bins are extracted from the residual signal. In the next phase, only the appropriate bin is selected, and the phase and amplitude features from this bin are passed through a learning algorithm to predict the expression of the subject's face as one of the four basic expressions: \textit{Sadness}, \textit{Happiness}, \textit{Surprise}, or \textit{Anger}. The following sections describe the individual modules of \ourmodel{} in detail.

\subsection{The Smartphone Application}
As shown in Figure~\ref{ver}, an Android phone usually has a camera on the top, besides an earpiece speaker. This speaker is responsible for transmitting in-call audio to the user. Apart from this speaker, every smartphone has a main speaker that facilitates the transmission of sounds like call alerts, music, etc. In addition, a typical smartphone generally contains one microphone at the bottom. This microphone aids in capturing voice and other ambient sounds. However, some modern smartphones are equipped with two or more microphones. In this work, we assess smartphones with minimal capabilities, thus considering only those with a single microphone. 

Figure~\ref{ver} shows the interface for data collection. We use the earpiece speaker for transmitting the chirps with low intensity. The main speaker can also be used for sending signals. The interface consists of simple play and stops buttons for starting and stopping the chirps. It also contains a text box for manually entering the ground truth label for expressions like \textit{Happiness}, \textit{Anger}, \textit{Surprise}, and \textit{Sadness}. Apart from that, we also use an automated ground-truth labeling method using image-based techniques. For this purpose, the application uses a camera preview that tracks the subject's facial region and automatically detects the expression using an image-based detection model. This camera-oriented feature has been used to validate the manually entered expression labels, as discussed in Section~\ref{basemodel}. Notably, we use the camera only to collect the ground-truth \revise{under a pure lab-scale controlled setup}, and the system does not need it during its actual runtime.   


\section{Design Details}\label{det}
We now discuss the details of each sub-module used in \ourmodel.
\subsection{Generation of FMCW Signals}
We consider FMCW signals or chirps that sweep from $f_{min}=16 $ kHz to $f_{max}=19 $ kHz. Each chirp is played for an optimal duration of $T=40$ ms; $T$ is directly proportional to the degree of overlap in the reflected echoes. Consecutive chirps are played while \ourmodel{} remains active and are separated by a silent period of $T_{sil}=30$ ms. $T_{sil}$ ensures that a recorded chirp does not contain a part of the next chirp being played in sequence, thus reducing the interference through overlap.

For a linear chirp, like the ones generated in \ourmodel, the instantaneous frequency ($f$) is dependent on the time ($t$) and can be expressed as,
\begin{equation}\label{freq}
f(t)=f_{min}+ct,
\end{equation}
where the chirp rate is $c = \frac{f_{max}-f_{min}}{T}$. The corresponding  phase of the same signal is an integration of $f(t)$ and can be expressed as, 
\begin{equation}\label{phase}
\phi(t)=\phi_{min}+2\pi(\frac{c}{2}t^2+f_{min}t),
\end{equation}
where $\phi_{min}$ is the phase at $t=0$. Now, by considering the sign of Equation~(\ref{phase}), the linear chirp can be expressed as the function of time $t$. Hence, 
\begin{equation}\label{funcT}
x(t)=sin[\phi_{min}+2\pi(\frac{c}{2}t^2+f_{min}t)]
\end{equation}

\subsubsection{Reception of Reflected Signals}
In exponential form, Equation~(\ref{funcT}) can also be written as,
\begin{equation}\label{funcT}
x(t)=e^{-j2\pi(f_{min}t+\frac{c}{2}t^2)}
\end{equation}
Let $r$ be the reflected signal that is also a function of time $t$. As discussed in Section~\ref{chirpchar}, a reflected signal is only a time-delayed ($t-\tau$) version of the original signal, having a time-of-flight $\tau$. Therefore, it can be expressed as,
\begin{equation}\label{refT}
r(t)=\sum_{p=1}^{N}\alpha_pe^{-j2\pi(f_{min}(t-\tau_p)+\frac{c}{2}(t-\tau_p)^2)}
\end{equation}
Notably, the reflected signal $r$ is a superimposition of multiple signals received from various environmental reflectors. For \ourmodel, the primary reflector is assumed to be the subject's facial region by default. However, besides reflectors like walls, furniture, etc., the echo also contains the direct path reflections from the speaker to the microphone. In Equation~(\ref{refT}), $N$ denotes the number of such multi-path signals. $\alpha$ is the signal attenuation. Similar to~\cite{10.1145/3463521}, we define the mixed signal to be recorded as a multiplication of the received signal with the complex conjugate of the transmitted signal, as follows. 
\begin{equation}\label{finT}
r_m(t)=\sum_{p=1}^{N}\alpha_pe^{-j2\pi(c\tau_pt+f_{min}\tau_p-\frac{c}{2}\tau_p^2)}
\end{equation}
The recorded signal is then passed through a high pass filter to allow the frequency range above $15.9$kHz. This automatically removes the ambient noises that fall within the audible frequency range.

\subsection{Processing of the Recorded Signals}
The in-device signal processing begins with synchronizing the speaker and the microphone. This allows us to remove the delay between the transmitted and the received signals caused by the device's imperfection. This is achieved by measuring the signal similarity between the generated and the reflected signals at different delays through cross-correlation. Assuming there are $N$ points in the transmitted and the reflected chirps, the normalized cross-correlation of the transmitted signal $x(t)$ and the received signal $r_m(t)$, shifted by $n$, can be expressed as,
\begin{equation}\label{xcor}
X_{corr}(n)=\frac{\frac{1}{N}\sum_{t=0}^{N}[r_m(t)-\overline{r_m}][x(t-n)-\overline{x}] }{\{\sum_{t=0}^N [r_m(t)-\overline{r_m}]^2\sum_{t=0}^N[x(t-n)-\overline{x}]^2\}^{\frac{1}{2}}}
\end{equation}
Here, $\overline{r_m}$ and $\overline{x}$ can be estimated by taking the average of $r_m$ over $N$ points and $x$ over $N$ points, respectively. Pertaining to the nature of the direct path signal, it should be maximally correlated to the generated chirp, as it suffers from no reflection. Thus, the delay at which the correlation value is maximum is considered in \ourmodel{} to synchronize the speaker and the microphone of the smartphone. Hilbert Transform~\cite{johansson1999hilbert} is used on the reflected signal to derive the analytic signal, i.e., its representation in the exponential form. As mentioned earlier, $r_m(t)$ is a product of this analytic signal and its complex conjugate in this domain. As we are interested in only the real part, the final expression of the mixed signal is just the real part of this complex multiplication.

This step is followed by the \textit{Fourier Transformation} of the received signal and static interference cancellation. In signal processing, static multi-path reflections are created from objects whose locations are fixed in the environment. For example, the chirp generated by the speaker will also be reflected from a wall, present behind the subject, a nearby table, and so on. Since these objects are static, the irrelevant noise induced by the reflections from these objects can be subtracted from the overall reflection capturing both static and dynamic reflections (caused by facial expressions). To achieve this, we first generate template recordings by transmitting chirps in an environment without the subject's presence. As a new session begins, where the subject is present in front of the device, the recorded template is subtracted from the reflected chirp, thus eliminating the external interference caused by static objects in the room.

The next step is to select the frequency bins that are most likely to capture the information about different facial regions (\textit{Facial AUs}). As we use a bandwidth of $3$ kHz, any two objects or points of reflection, separated by a distance of $5.6$ cm or more, will result in a reflected echo, having distinct frequencies. For example, while the signal path generated by the reflection of the transmitted chirp from the eye region will fall in a frequency bin $f_1$, that from a wall behind will fall in a different frequency bin $f_2$. To further ensure that the frequency bin capturing the information about the face is selected, we depend on the phase-amplitude variation induced by AUs. Thus, we choose the frequency bin with the maximum variance, as shown in~\cite{10.1145/3463521}. 

\subsection{Prediction of expressions}
Finally, the phase and amplitude of the signal with the selected frequency are used to predict the subject's facial expression using an ensemble of three different classifiers using a majority voting technique. The three classifiers are chosen by comparing the classification accuracy of different algorithms. \revise{ Empirically, we observed that the average accuracy of \textit{Support Vector Machine} was 17.5\%, \textit{3-Nearest Neighbour} was 44.1\%, \textit{Adaboost with $50$ estimators} was 47\%, \textit{MLP Classifiers} was 49\%, \textit{Random Forest with maximum depth of 10} was 72.72\%, \textit{Decision Tree} was 96.61\%, \textit{Logistic Regression} was 66.41\% and \textit{Naive Baye's Classifier} was 36.29\%. In our implementation, the three classifiers with the highest accuracy, which were chosen for the ensemble, were -- Logistic Regression with L2 Penalty \footnote{\url{https://medium.com/@aditya97p/l1-and-l2-regularization-237438a9caa6} (Accessed: \today)} and Limited-memory Broyden–Fletcher–Goldfarb–Shanno (BFGS) solver\footnote{\url{https://towardsdatascience.com/limited-memory-broyden-fletcher-goldfarb-shanno-algorithm-in-ml-net-118dec066ba} (Accessed: \today)}, Decision Tree with Gini impurity \footnote{\url{https://towardsdatascience.com/gini-impurity-measure-dbd3878ead33} (Accessed: \today)} and the Random Forest with a maximum depth of $10$.} By using the majority voting method with these three best classifiers, we achieved an overall improvement of $\sim4\%$ in the classification accuracy, as compared to the average accuracy of the individual models (details in Section~\ref{basemodel}). 

\section{Implementation, Resource Profiling, and Evaluation Methodology}
\label{sec:eval}
\revise{This section provides the implementation details and resource consumption benchmarking of \ourmodel{} and an overall discussion on how we conduct the evaluation of the proposed system in a principled way. The implementation of \ourmodel{} along with partial data (anonymized) has been made open-sourced\footnote{Code link: \url{https://github.com/anonymous0304/ExpresSense.git} (The link has been anonymized for double-blind review).}.} 

\subsection{Implementation Apparatus}
\ourmodel{} has been developed for Android Platforms, using Android Studio\footnote{\url{https://developer.android.com/studio} (Accessed: \today)}. A sampling rate of 44100 Hz has been considered for the FMCW signals, which are encoded using Pulse Code Modulation (PCM)  to represent the sampled signals. For transmitting these signals, the \texttt{AudioTrack}\footnote{\url{https://developer.android.com/reference/android/media/AudioTrack} (Accessed: \today)} class has been used, which allows the streaming of Pulse Coded signals. In order to observe the general capability of commodity smartphones in  facilitating acoustic sensing-based facial expression detection, we have installed and used the application on different commodity smartphones (Realme and Samsung) having 4, 6 and 8 GB of RAM. The minimum Android Version considered is 9. The chipset of the tested devices include Qualcomm Snapdragon 730G, Qualcomm Snapdragon 710, Qualcomm SDM730 Snapdragon 730G (8 nm), MediaTek Helio G95 (12 nm), Qualcomm SDM675 Snapdragon 675 (11 nm), and so on. For testing the accuracy of different models in predicting expressions, we train and test the various models offline using Python sckit-learn\footnote{\url{https://scikit-learn.org/stable/} (Accessed: \today)} library.  Finally, the trained model is uploaded to the Heroku\footnote{\url{https://www.heroku.com/} (Accessed: \today)} platform and connected to the smartphone application using an intermediate FLASK-API\footnote{\url{https://flask.palletsprojects.com/en/2.1.x/} (Accessed: \today)}. It is to be noted that the entire signal processing part is executed locally in the user's smartphone and only the numeric values of generated amplitude and phase are sent to the remote server for being predicted. The class label is then communicated back to the user's device. For FMCW signals, range resolution can be defined as the  ratio between the speed of sound in the air and twice the bandwidth of the signal~\cite{10.1145/3463521}. Hence, for a signal with a bandwidth of $3$ kHz, if there are two reflected signals, caused by two objects, placed at a distance of $5.6$cm or more, with respect to the sound source, then these two signals will fall in different frequency bins. Hence, we need to select a particular frequency that corresponds to the reflection from the facial region of the person. Due to some of the inherent limitations of a smartphone (discussed in Section \ref{lim}), it is difficult to reduce the range resolution of the signal, without making it overlap with the audible sounds. Thus, we assume that the user's face and any other nearby object are separated at least by a distance of $5.6$ cm. If there is an object very close to the face ($<5.6$ cm), then the selected frequency bin will also reflect the information of the second object, along with the face. 

\subsection{Profiling the Resource Consumption}
\ourmodel{} consumes around $75$-$112$ MB RAM during the runtime. The lightweight nature of our model is tested by fully charging a smartphone and keeping the application on till the charge drops to 20\%. The application could continuously run for more than $7$ hours. Further, to ensure near real-time processing, we analyzed the time taken to process the received signals using different smartphones with $4$, $6$, and $8$ GB RAM. \revise{For this purpose, we calculate the processing time for each chirp for different expressions. The time reflects the total time in which a reflected chirp is captured through the microphone and processed locally in the smartphone to generate the amplitude and phase values from the selected frequency bin.} The average processing time was estimated to be $\approx5$ seconds, $\approx3.5$ seconds, and $\approx1$ second, respectively, with the three different RAM availability. 

\subsection{Evaluation Methodology}
\revise{To evaluate \ourmodel{} in a principled manner, we set the following objectives to analyze the system thoroughly under different aspects.} 
\begin{enumerate}
    \item \revise{How well can \ourmodel{} infer the four basic facial expressions of a subject in general?}
    \item \revise{How do different environmental factors, like the elevation, orientation, and tilting of the phone, motion of the subject, ambient sound, hand placement, glasses, finger movement, etc., impact the performance of \ourmodel?}
    \item \revise{How does \ourmodel{} perform under natural expressions?}
    \item \revise{How usable \ourmodel{} is in practice?}
\end{enumerate}

\revise{Evaluating \ourmodel{} under objectives (1) and (2) above is straightforward, as we can go for a controlled lab-scale setup where trained subjects can pose for different expressions under different conditions for a short and fixed duration while holding the phone in front. We can also collect the ground truth using other modalities, such as self-annotation, annotation through one or more dedicated volunteers, or well-established vision-based automated labeling techniques by capturing the subject's face through the phone's front camera. We performed controlled experiments to evaluate \ourmodel{} in a general setup, as discussed in Section~\ref{basemodel}.} 

\revise{However, evaluating \ourmodel{} under Objective (3) is challenging. First, we need third-party applications that can naturally trigger changes in the subject's facial expression. For example, a video streaming app may trigger natural changes in facial expression based on the genre of the video being streamed. Second, annotating the data is challenging as the facial expressions may change continuously, so we need precise time boundaries when the expression changes. Human annotation cannot work with this precision. Further, automated annotation using the camera may cause discomfort to the subject or may divert their attention, thus affecting their natural expressions. To solve this issue, we use an indirect way of evaluating the system by utilizing the existing research on gauging human engagement through facial expressions~\cite{dubbaka2020detecting,bosch2016using,sharma2020assessing}. We match the temporal changes of the subject's facial expression with the video genre and derive an engagement score for the entire duration of the video streaming session. We then collect the ground truth through questionnaires and self-assessment and match the ground truth with the computed engagement score. The underlying hypothesis is that if there is a good match between the computed and the ground-truth engagement scores, then \ourmodel{} has captured the facial expressions accurately. Section~\ref{sec:natural} discusses this evaluation methodology and the corresponding results in detail.}

\revise{Finally, we performed a thorough usability study in the wild using the proof of concept (PoC) video streaming application that can inform the subjects about their engagement level while watching the video. The details have been discussed in Section~\ref{usability}.}

\section{Evaluating \ourmodel{} under a Lab-Scale Controlled Environment}
\label{basemodel}
First, we explore the performance of \ourmodel, as an individual module for detecting facial expressions based on acoustic chirps from a standalone smartphone. The details follow. 

\subsection{Experimental Setup}\label{coreMeth}
The data collected for training and testing \ourmodel{} has been generated in a monitored setup from  $10$ participants (P1-P10) who volunteered in the evaluation. These $10$ participants ($4$ females, $6$ males) belonged to different age groups and professional backgrounds. Two participants belonged to the age group of $20$-$25$ years, four participants belonged to the age group of $26$-$49$ years, and the rest belonged to the age group of $50$-$65$ years. To ensure  professional diversity, we chose the participants in such a way that three belonged to the IT industry and were Software Engineers, two were Undergraduate students, two were home tutors, one belonged to the banking sector, one was a research scholar, and one was a retired personnel. Four of them used glasses and others had normal eyesight.

In this setup, the participants were asked to place the smartphone at a distance of $\approx30$ cm from their faces. The angle of elevation of the device with respect to the face was not fixed a priori; however, in all the cases, the participants preferred to place the phone at about  $-20^o$ (on the vertical axis), corresponding to the face. The azimuth angle of the smartphone with respect to the face was roughly $0^o$ (directly in front of the face), as this was the natural viewing angle for all the users. The subjects were asked to hold the smartphone by hand or use a smartphone holder for convenience during the session. One of the participants performed the experiment in complete darkness, while others performed under normal lighting conditions. The experiment was performed indoors, in the presence of natural ambient sounds generated by ceiling fans, outdoor noises (like cars passing by, children playing on the ground, etc.), and so on. For collecting the data, the methodology aligned with that explained in Section~\ref{meth}. For each participant, the entire experiment was conducted in three different sessions, preceded by a training session. In each session, the participants reproduced the four facial expressions in different orders. In each session, every expression was captured for a duration of $1$ min when the participant could render different variants of a chosen expression, followed by a pause between two expressions. The participants were free to choose the pause duration. We advised the participants to keep the expressions as natural as possible.

\subsection{Ground Truth}
The ground truth generation and validation has been conducted in a $3$-layered technique. Firstly, the data was labeled manually by the participants themselves. During the pause phase between two expressions, the participants entered the expression to be produced next in a text box in the application. These expression labels acted as the ground truth. Secondly, automated labels were generated. The application constantly monitored the facial expressions of the subjects through the camera and automatically predicted the expressions using the trained MobileNet model \cite{sandler2018mobilenetv2}, incorporated into the application. For the automatic detection of the facial expressions, we have used the MMA Facial Expression dataset\footnote{\url{https://www.kaggle.com/datasets/mahmoudima/mma-facial-expression} (Access: \today)} which contains about $92,958$ training  and $17,356$ testing images from different expression categories like surprise, fear, angry, neutral, sad, disgust, happy. The MobileNet V2~\cite{sandler2018mobilenetv2} model has been trained for this purpose as it provides a significant accuracy (loss=.01) and performs real-time predictions on mobile devices. Thirdly, labels were verified through close monitoring. During the experiment, the participant's expressions were closely monitored by two different individuals to ensure that the correct expression is being produced with respect to the manually entered label and the expression sequence. In case of disagreement, the participants were requested to repeat the expression.

Finally, the manually entered and auto-generated labels were synchronized and compared for validated sessions. It is to be noted that, pertaining to the possibility of misclassification of expressions by MobileNet, more weight was given to the manually entered labels. For example, if for one minute of ``\textit{Happy}'' (manually entered) expression, the MobileNet predicted expression was ``\textit{Happy}'' for the majority of the data, then the rest of the data labels, although classified as a different expression, were also considered as ``\textit{Happy}''. Finally, a simple pruning was performed to select the data points generated about $3$--$5$ seconds after clicking the ``\textit{Start Chirp}'' button and those generated before $3$--$5$ seconds of clicking the ``\textit{Stop Chirp}'' button. This is because we observed that for most of the participants, the actual expressions started a little after starting the chirps as they shifted their focus from clicking the button to creating the expression during that time. Similarly, the participants released the expressions slightly before stopping the chirps as they cognitively prepared to press the button. 

\revise{\textbf{Dataset collected.} From each session, we finally collect a total of $20$ data points (each of $\approx1$ min duration with labeling and pruning as discussed above) for each class in a session, i.e., a total of $80$ data points per session. Each data sample is a pair of amplitude-feature values, along with the ground truth label. The entire dataset thus contains $\approx2400$ data samples from all four classes across all $10$ participants combined. Apart from these regular sessions, the users were also asked to attend additional sessions where data was created in a similar method for different conditions like elevation and angular change of the device, different ambient sound levels, degree of motion,  different hand positions for holding the device, \revise{different degree of finger movement, and presence of surrounding objects}. In total we have collected $\approx 6880$ data samples for the lab-scale controlled experiments, which have been divided into train-test splits for different test scenarios, as we explain later for individual cases.}

\subsection{Results}
We next discuss the user-specific system performance of \ourmodel{} and its sensitivity analysis. 

\subsubsection{User-Specific System Performance}
We first explore the performance of \ourmodel{} for individual subjects. We analyze \ourmodel{} from three perspectives -- \textit{overall performance}, \textit{inter-session performance} and \textit{intra-session performance}. In the overall performance testing, we mix the data from all sessions and make a 4:1 data segmentation for training and testing the model based on stratified random sampling. By shuffling the data and randomizing the segmentation, we run the learning model $10$ times and present the average accuracy of the system per subject. To further analyze whether the system can capture the characteristics of the subjects from independent sessions, we perform an inter-session performance estimation. Here, we train the model using data from two sessions and predict the data derived from the third experimental session of individual subjects. This allows us to better estimate if \ourmodel{} can capture the overall characteristics of a subject from independent sessions. \revise{Finally, in the intra-session study, we trained the model with $\approx16$ out of the $20$ data points for each expression class in a single session, and the rest $4$ data points per class per session were used to test the model.} 

\begin{figure}[!ht]
\begin{minipage}[b]{0.45\textwidth}
   \centering\includegraphics[width=\textwidth]{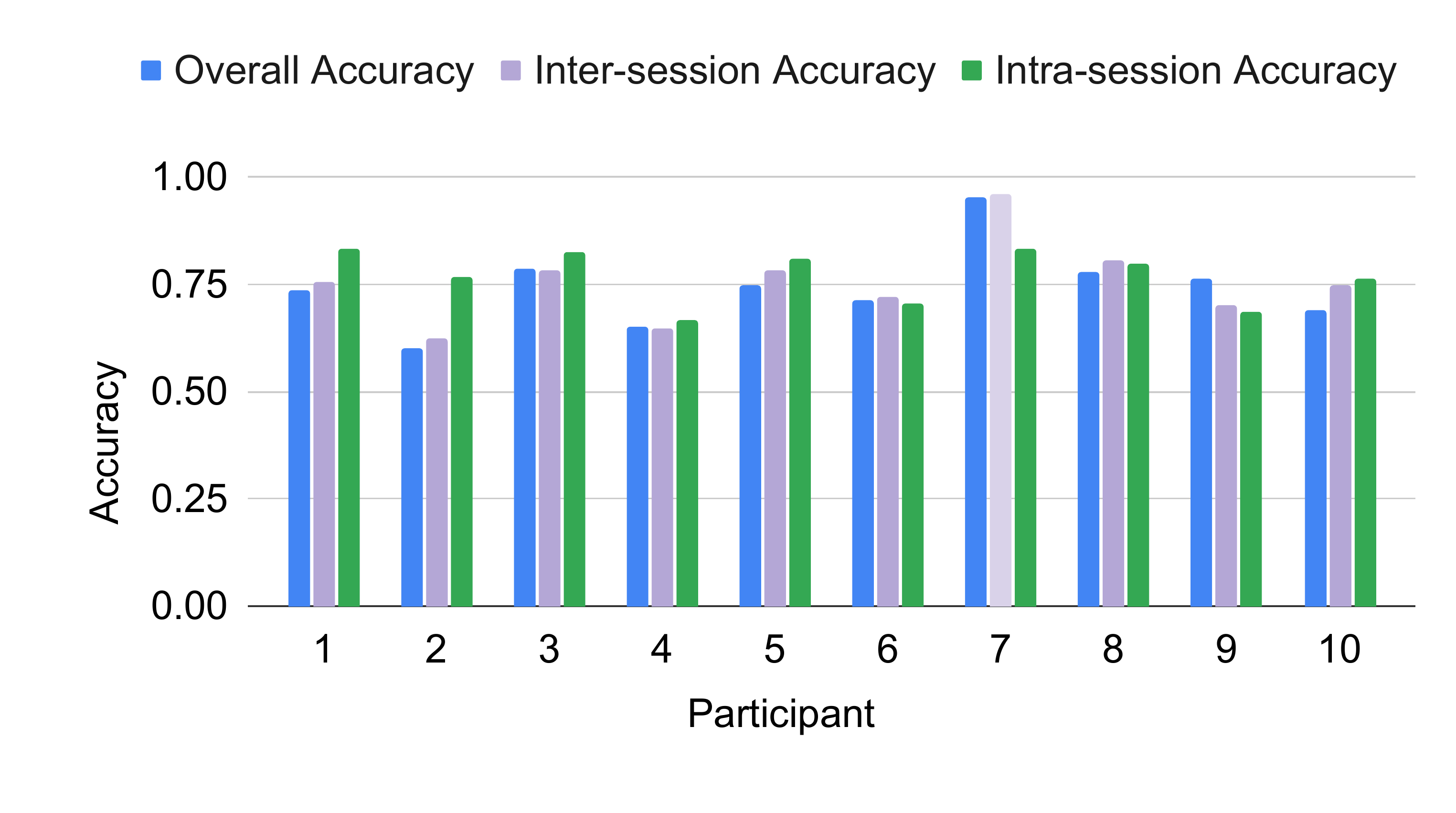}
  \caption{Comparison of participant-wise variation  of overall, inter-session and intra-session accuracy}
  \Description{A bar plot showing that the intra and inter-session accuracy for  each participant is higher than the overall accuracy. }
  \label{compac}
\end{minipage}%
\hfill
\begin{minipage}[b]{0.45\textwidth}
 \centering \includegraphics[width=\textwidth]{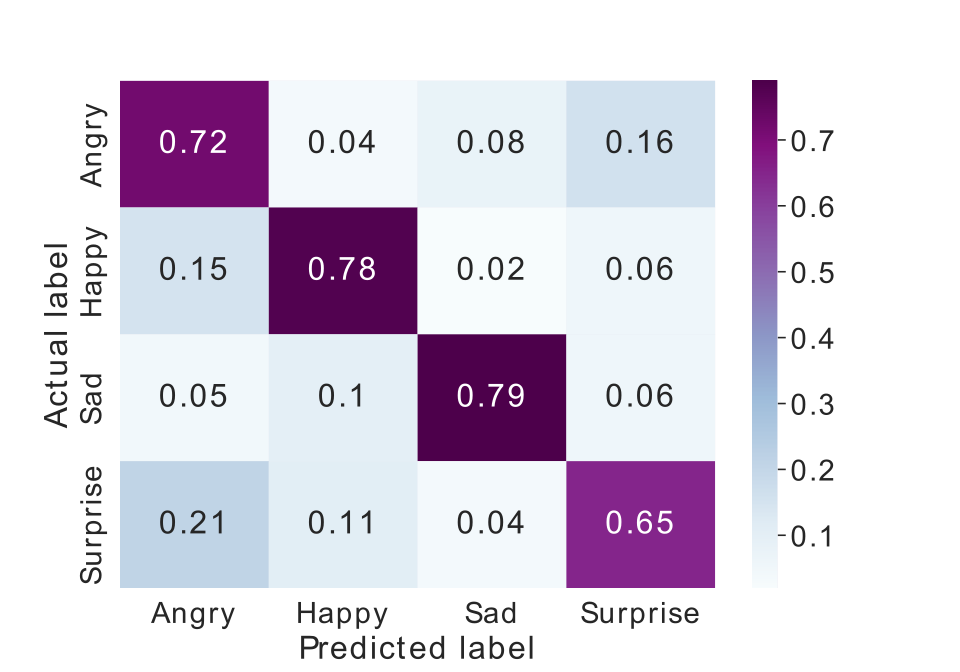}
  \caption{Overall classification accuracy of individual expressions}
  \Description{A confusion matrix for  classification of individual expressions by ExpresSense-Core.}
  \label{confusionOverview}
\end{minipage}
  
\end{figure}

Figure~\ref{compac} shows the comparison of the overall, inter-session, and intra-session accuracy of detecting expressions of individual subjects. In the \textit{overall performance} assessment, \ourmodel{} achieves an average accuracy of $\sim73\%$ across all the subjects. In \textit{inter-session performance} testing, we achieve an average accuracy of $\sim73.5\%$. However, it can be seen that the individual accuracy for some subjects from the inter-session study has been improved ($\approx 2.45\%$) than the overall accuracy (when data is shuffled). This infers that the model is able to learn the feature pattern better when individual sessions are fed to it, particularly because of the inter-session variations caused by the dislocation of positions and renewed expressions.

In \textit{intra-session performance} testing, we achieve an average accuracy of $74.6\%$. The observation aligns with the expected output, as individual sessions are more likely to have distinctive features due to the lack of involuntary positional and angular shifts. However, for P7, the accuracy dropped from $96\%$ in inter-session evaluation to $83\%$ in the intra-session study. Notably, P7 took less pause time than other participants, which induced muscular fatigue, causing the participant to rapidly produce variations while holding individual expressions.

\revise{To further analyze the performance of \ourmodel{} in classifying individual expressions, Figure \ref{confusionOverview} shows the confusion matrix for the \textit{overall performance} of the system. In this figure, the true positives, true negatives, false positives, and false negatives are derived by considering the average detection results of the $10$ subjects. The figure depicts that the highest accuracy has been achieved for the \textit{``Sad''} expressions, closely followed by \textit{``Happiness''}. Although the accuracy for the class \textit{``Angry''} is comparable to these classes, that of the class \textit{``Surprise''} is less than the rest. The matrix shows that this expression has mostly been confused with \textit{``Angry''} and vice versa. By considering the AUs that generate expressions like \textit{``Anger''} and \textit{``Surprise''}, we observe that both these expressions have overlapping characteristics like widening of eyes and furrowed brows (similar characteristics of ocular AU) along with distinct characteristics like tensed mouth, jaws in anger and relaxed or dropped jaws in surprise (dissimilar characteristics of oral AU). However, depending on the individual, the expression of anger can also show similarities in the characteristics of the oral AU with that of surprise. This explains the performance of the system in detecting \textit{``Surprise''} with lower accuracy as the analysis of the underlying characteristics of AUs aligns with the observation (refer to Section \ref{pilorobs}) that ocular AUs have the highest effect on the signal features. Similarly, the overlaps between other expressions have resulted from the similarity in different AUs (e.g., partial visibility of teeth for \textit{``Happy''} and \textit{``Angry''}).  } 

\subsection{Sensitivity Analysis}\label{sensi}
\revise{We next analyze the sensitivity of \ourmodel{} under different environmental conditions that may impact the performance of the system.}

\begin{figure}[!ht]
    \subfloat[Distance]{\includegraphics[width=0.3\textwidth]{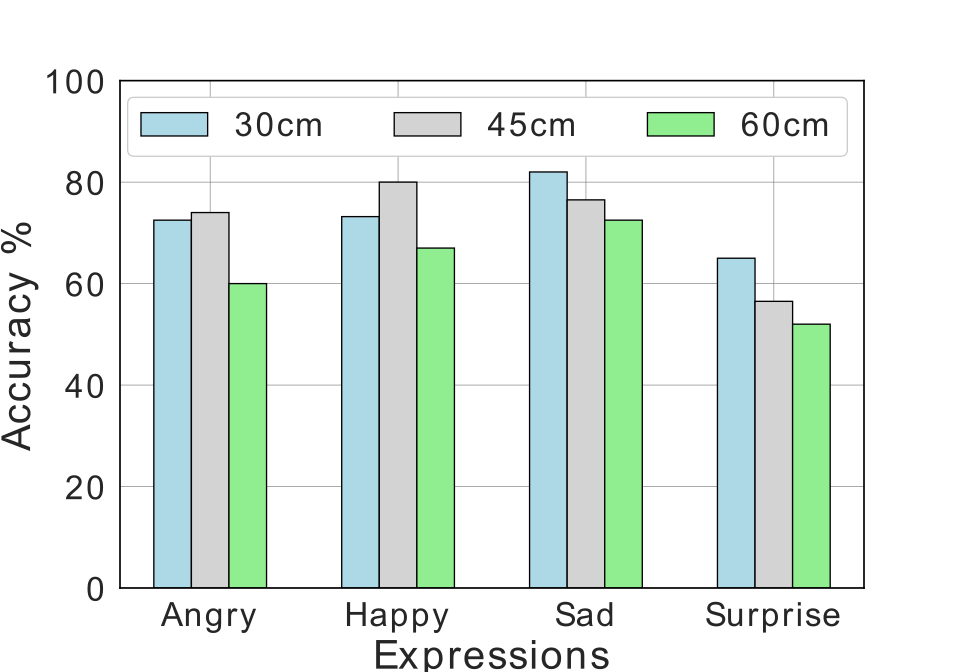}\label{db}}
    \subfloat[Elevation]{\includegraphics[width=0.3\textwidth]{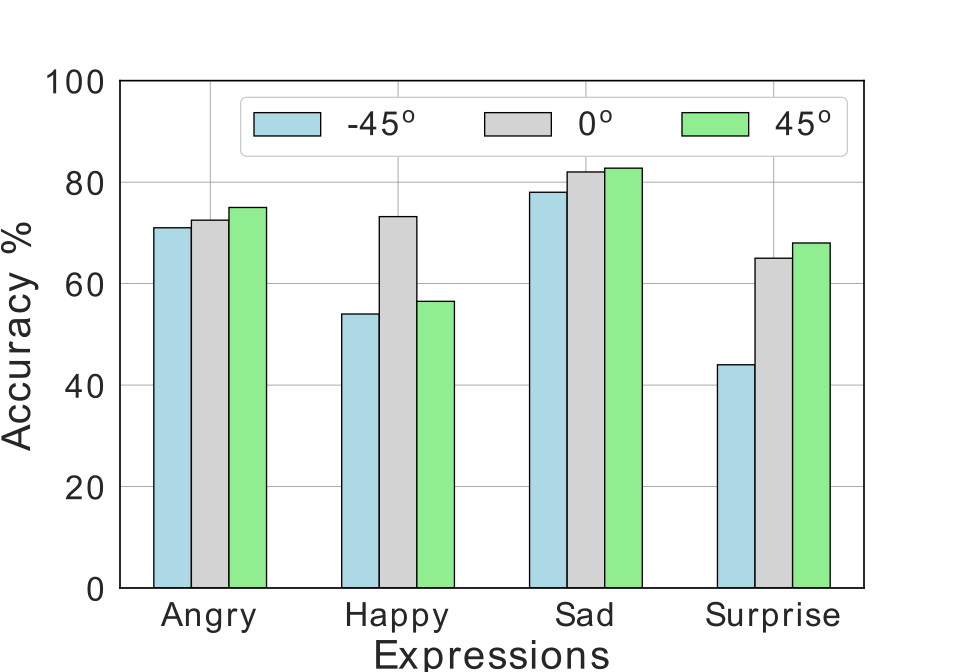}\label{eb}}
    \subfloat[Tilting Angle]{\includegraphics[width=0.3\textwidth]{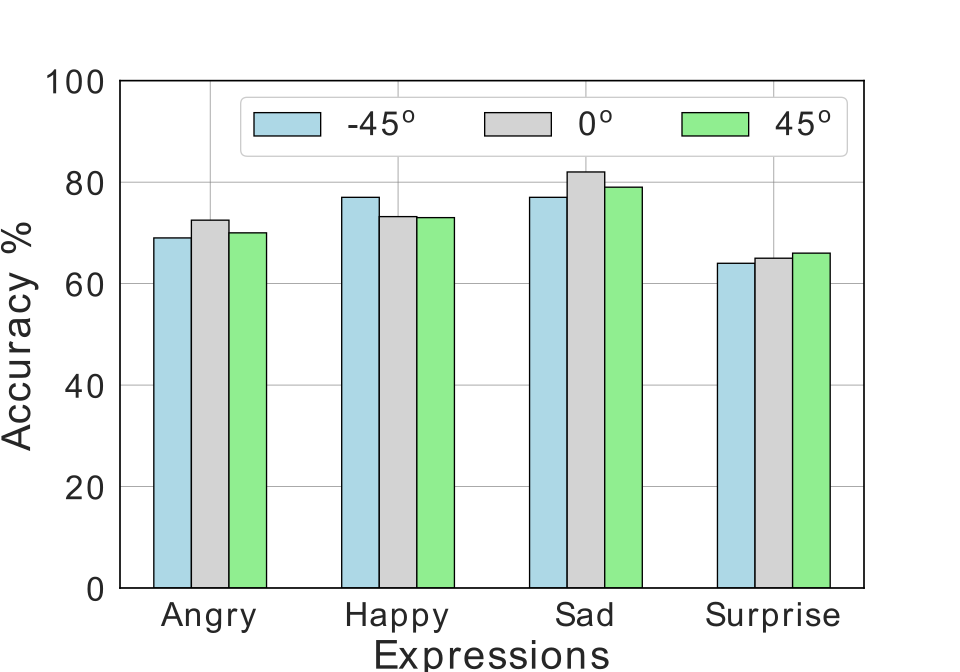}\label{angb}}\\
    \subfloat[Ambient Sound]{\includegraphics[width=0.3\textwidth]{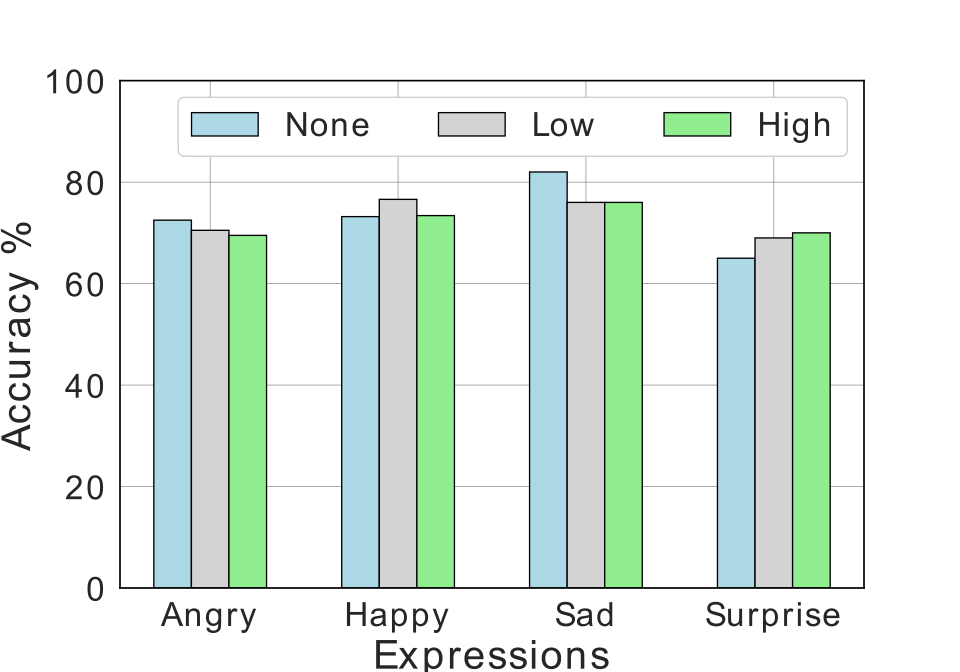}\label{sb}}
    \subfloat[Motion]{\includegraphics[width=0.3\textwidth]{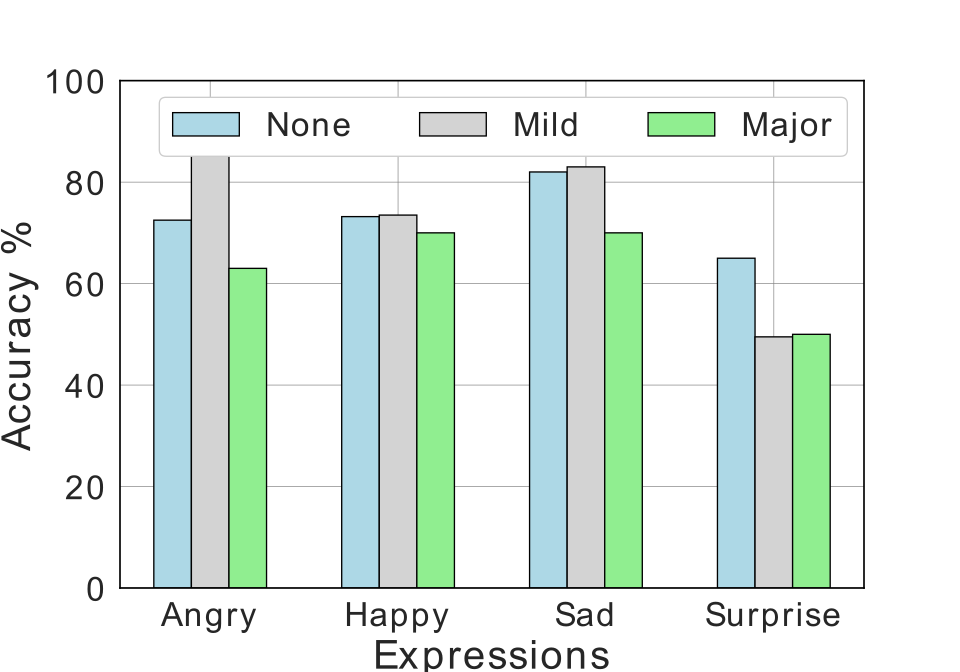}\label{mb}}
    \subfloat[Hand Placement]{\includegraphics[width=0.3\textwidth]{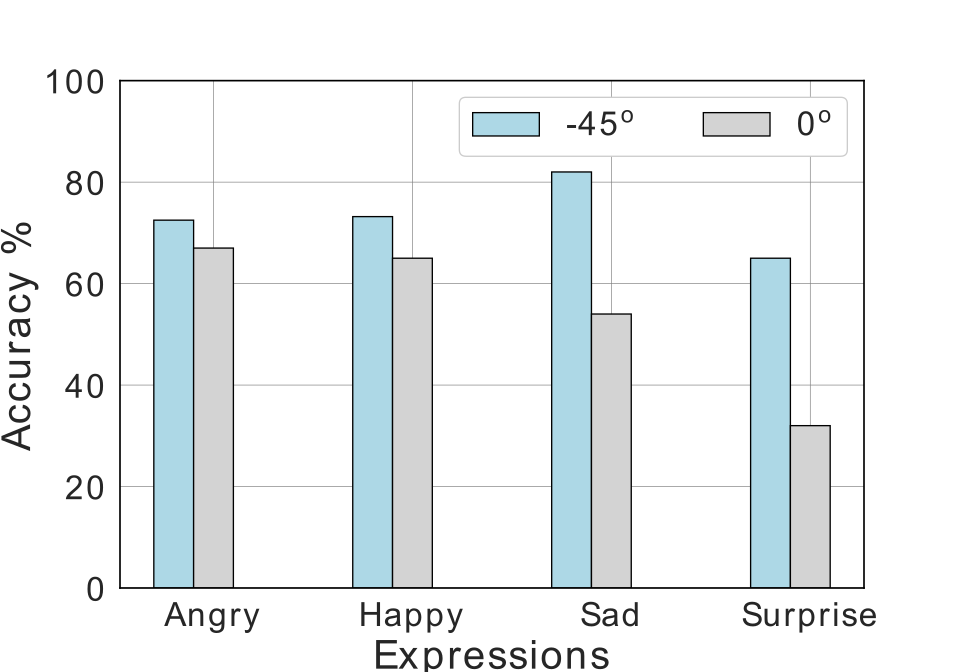}\label{hpb}}\\
    \subfloat[Finger Movement]{\includegraphics[width=0.3\textwidth]{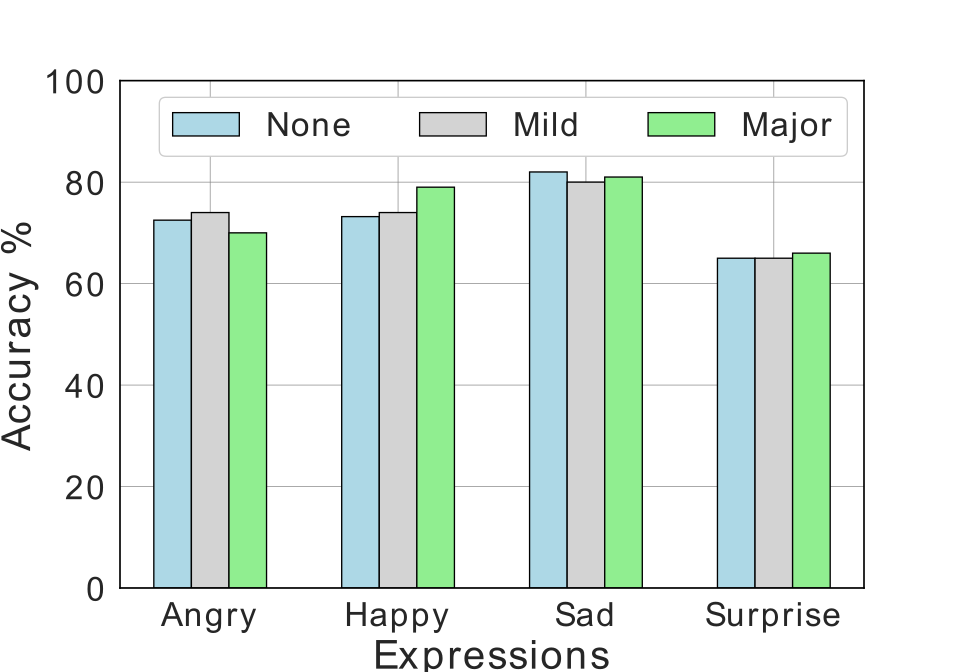}\label{fib}}
    \subfloat[Reflective Surface]{\includegraphics[width=0.3\textwidth]{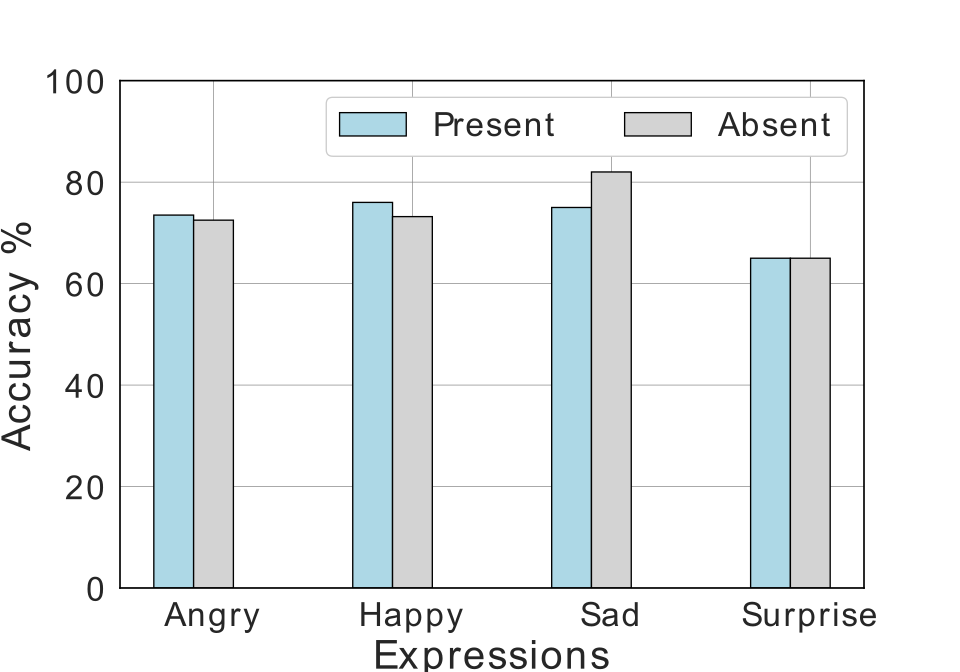}\label{monb}}
    \subfloat[Glasses]{\includegraphics[width=0.3\textwidth]{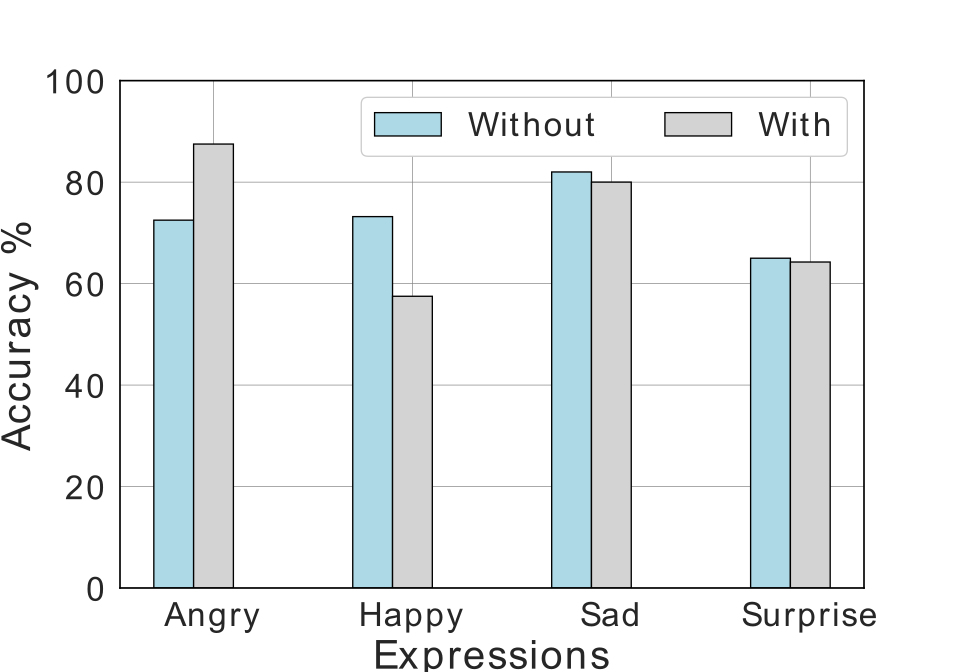}\label{gb}}
    \caption{\revise{Sensitivity Analysis of \ourmodel{}: Impact of Various Environmental Factors}}
    \Description{A group of nine bar charts, showing the accuracy of each expression under different experimental setups like face to device distances, elevations, tilting angles, ambient sound, motion, hand placement while holding the phone, finger movement, presence of reflective surface and eye glasses. }

    \label{fig:sensitivity}
\end{figure}

\subsubsection{Impact of Distance}
To analyze the effect of distance on the classification accuracy, we asked the subjects to place the smartphone at $30$ cm, $45$ cm, and $60$ cm from their face, in three consecutive sessions, respectively. While recording the reflected signal from three different distances, the devices were placed at zero-degree elevation and tilting angles to the user. Figure~\ref{db} shows that the performance of \ourmodel{} drops ($\approx6$-$14\%$) as the distance is increased. \revise{Notably, we observe maximum drop in the performance ($\approx 14\%$) for the \textit{``Surprise''} class as it gets confused with \textit{``Angry''}.}


\subsubsection{Impact of Phone's Elevation}
In this study, the distance between the subject and device was kept fixed at $30$cm, and the angle of elevation (vertical height) of the device was varied from $-45^{\circ}$ to $+45^{\circ}$ with respect to the subject's face. Interestingly, even though the performance of the system (Figure~\ref{eb}) was comparable ($74\%$ and $72\%$, respectively) for $0^{\circ}$ and $+45^{\circ}$ of device's elevation, at $-45^{\circ}$, the accuracy of the system dropped to $62\%$. The contributing factor behind this observation was the location of the microphone. At $-45^{\circ}$, the microphone of the smartphone, placed at the bottom of the phone, captured a noisy signal due to interference from the parts of the upper body.

\subsubsection{Impact of Phone's Tilting Angle}
Similar to the previous study, in this analysis, the devices were kept at $-45^{\circ}$ (left), $0^{\circ}$ (in front), and $+45^{\circ}$ (right) horizontally, with respect to the subject's face, at an elevation of $0^{\circ}$. However, in this case, the angle had no significant impact on the results (Figure~\ref{angb}).  


\subsubsection{Impact of Environmental Noise}
\ourmodel{} uses near-ultrasound signals that should not interfere with most of the audible frequency range. To test this hypothesis, we asked the subjects to produce facial expressions in three different environments, while the smartphone with \ourmodel{} was placed at $0^{\circ}$ of vertical and horizontal angles from the user's face at a distance of $30$ cm. In the first case, we aimed to eliminate all possible sources of sound. It is to be noted that complete silence ($0$ dBa) is not possible to attain, even in a lab-scaled study. Hence, by no sound, we indicate the absence of all audible ambient noises created by ceiling fans, keyboards, etc. In the second level, we induced noise between $\sim15$-$30$ dBa, which was generated by human whispers, ceiling fans, and so on. This was marked as an environment with low ambient sound. In the third level, we incorporated a high sound level by playing background music, loud conversations, and traffic sound. \revise{ Figure~\ref{sb} shows that environmental sounds did not significantly affect the overall and class-wise accuracy of the system. This is because most of the environmental sounds fall well below the frequency of $16$ kHz and are eliminated by the high pass filter of \ourmodel{}. However, the performance of the system will be adversely affected if ultrasound signals that overlap with the frequency range of the chirps in \ourmodel{} are introduced in the environment (refer to Section \ref{lim}).}

\subsubsection{Impact of Motion}
Next, we assess the effect of motion for three scenarios -- (1) \textit{No motion}, where the smartphone was placed on a phone holder, (2) \textit{Mild motion}, when the phone was held by a hand, while the users remained seated and (3) \textit{Major movement} created by allowing the users to walk in the room while holding the phone by hand. Figure~\ref{mb} shows that only major movements decrease the system's accuracy by $\approx9\%$ on average. \revise{Further, it had the most significant effect on the detection accuracy of the expression \textit{``Angry''} (average decrease of $18.2$\%) and \textit{``Sad''} (average decrease of $12.5$\%). The effect of major motion can be explained by the workflow of \ourmodel{} where bin-selection (calibration) is a one-time process. Any large change in the position and distance of the reflector (face), as indicated by the selected bin, caused by body motion or change of hands will cause the system's accuracy to be affected. However, this can be solved by re-calibrating the system as and when the smartphone's inertial sensors detect large body movements.}
 
\subsubsection{ Effect of Hand Placement}\label{handp}
Next, we estimate if the holding position of the phone affects the system's performance. For this study, we asked the subjects to (1) place their fingers on the side of the phone and (2) place their palm toward the bottom of the phone. Figure \ref{hpb} shows that placing the fingers on the side allows the system to perform with an accuracy of about $74\%$ while placing the palm towards the bottom decreases the accuracy to $55\%$. This is because, in the latter case, the palms cover the microphone, making it unable to capture the reflected signals completely. 

\revise{\subsubsection{Effect of Finger Motion}\label{fingm}
To analyze whether the system is affected by the movement of fingers, we asked the users to (1) restrict the movement of fingers, (2) periodically reply to text messages received in a floating window on the smartphone's screen (requires gentle finger movement), and (3) continuously chat or use the video controls (requires significant finger movement) while using the system in the background. Figure~\ref{fib} shows that there was no significant effect of finger movement on the overall and expression-wise accuracy of the \ourmodel{}. This observation was fascinating as it did not align with the impact of (body) motion on the system's performance. However, it should be noted that the reflector's position (face) was fixed for this experiment, and the calibration phase captured the presence of the fingers that were kept static during calibration (which takes about $4$ seconds). However, if the fingers are moved during the calibration phase, the bin selection can be affected, thus decreasing the overall system performance. This overhead can be eliminated by detecting the degree of change in the signal's amplitude and phase, as a movement in facial AUs is significantly more fine-grained than finger movements. }

\revise{\subsubsection{ Effect of Surrounding Objects}\label{monitors}
In this experiment, the subjects were asked to perform the same experiment while (1) a monitor (a reflective surface) was placed at about $45$ cm behind the smartphone, and (2) no object was present within a distance of $60$ cm from the smartphone. This experiment aimed to test whether surrounding reflectors like a monitor can affect the system's performance. Interestingly, these static objects did not affect the system's performance (Figure~\ref{monb}). This observation can be explained through the fact that static interference cancellation is performed by \ourmodel{} in the signal processing stage.}

\revise{\subsubsection{Effect of Glasses}\label{glassw}
Finally, we present the result by considering the subjects who (1) wore glasses and (2) did not wear glasses. Figure~\ref{gb} shows that the performance of \ourmodel{} is not affected by the presence of glasses.}

\section{Evaluating \ourmodel{} Under Natural Expressions}
\label{sec:natural}
\begin{wrapfigure}{l}{0.4\textwidth}
\includegraphics[width=.4\textwidth]{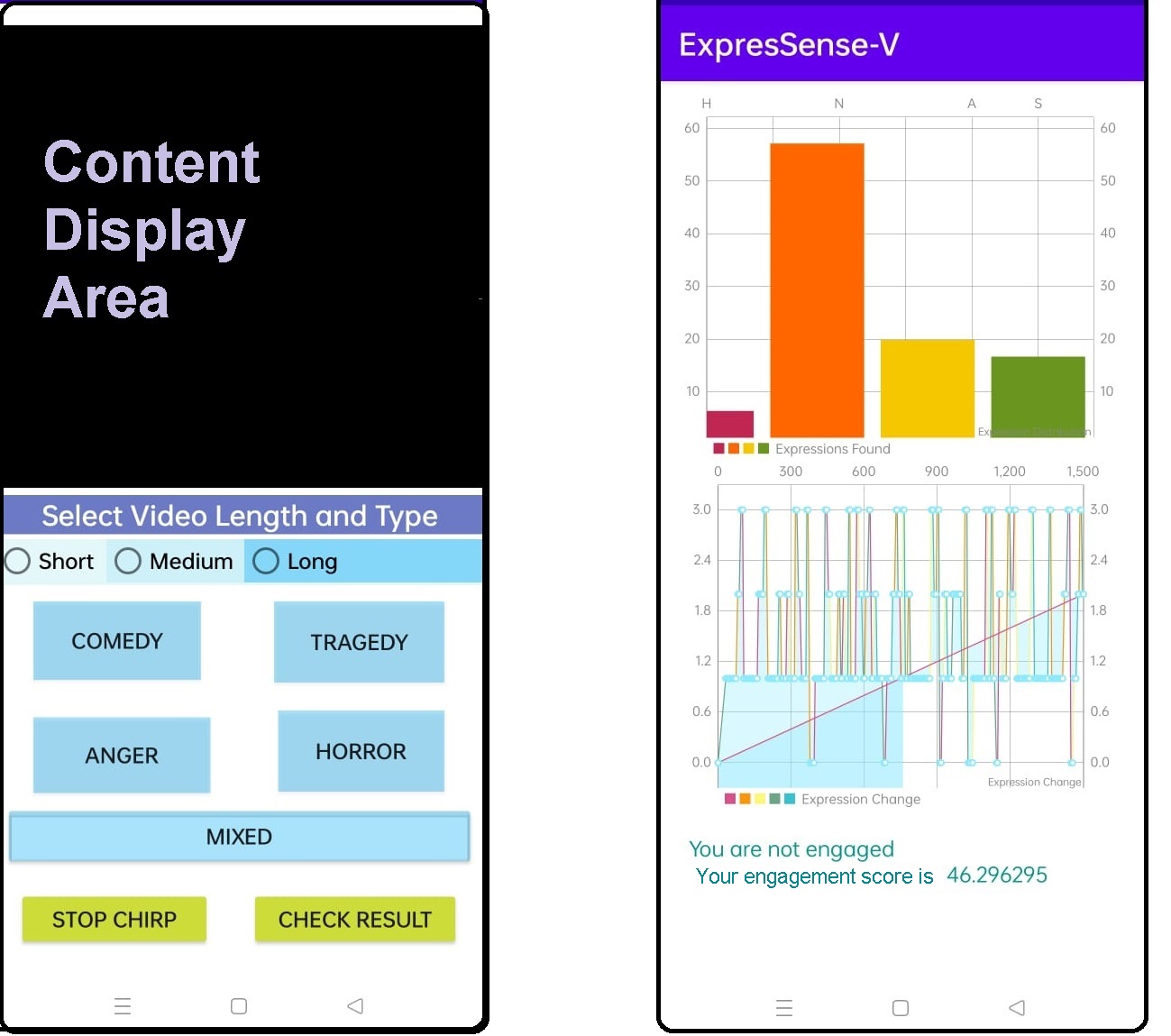}
\caption{Interface of \ourmodel{} Video Streaming App: The Content View (left) and the Result View (right)}
 \Description{The figure shows two interfaces of ExpresSense   video streaming app: the left interface has a content display area, three radio buttons for video length selection, five buttons labeled as comedy, tragedy, anger, horror and mixed. It also has a stop chirp button and a check result button. The second interface shows two expression graphs (bar and line) and the engagement indicator and engagement scores.}
\label{app}
\end{wrapfigure}
As discussed in Section~\ref{sec:eval}, we developed a smartphone video streaming app that uses \ourmodel{} at its core to continuously sense the facial expressions of the subject while watching a video and measure how engaged the subject was during the session. The application has been open-sourced\footnote{Code link: \url{https://github.com/anonymous0304/ExpresSense.git}} so that participants can use it in an uncontrolled environment and record their feedback. 


Figure \ref{app} depicts the interface for \ourmodel{} video streaming app. The image to the left displays the \textbf{Content Display Area} where the user can view YouTube Videos. The user can also select the \textbf{length of the content} as either short (7-10 mins ), medium ($\approx$15 mins), or long ($\approx$30 mins). Moreover, the user is allowed to select the \textbf{genre} of the content as either: \textit{comedy}, \textit{tragedy}, \textit{horror}, \textit{anger}, or \textit{mixed}. Once the user selects these fields, they can click on the \textbf{start chirp} button, which starts the video and simultaneously emits the chirp signals as described in \ourmodel. The received signals are processed using the pipeline mentioned in Section~\ref{det}, and the final phase and amplitude values are transmitted to the trained ensemble model to predict the expression label. Thus, the application continuously monitors the instantaneous expressions and estimates the rate of expression change throughout the video. As the user completes watching the content, they can click on the \textbf{Check Result} button, which opens the interface shown to the right of Figure \ref{app}. In the result view, the user can see the percentage of each expression, as predicted by \ourmodel{} throughout the usage of the application. Along with this, the user can also see how their expressions varied over time. These two graphs provide the user with a visual representation of their engagement which is then summarized as detected \revise{engagement indicator} and the overall engagement score. We use a rule-based approach, as discussed next, to generate the \revise{engagement indicator} and engagement scores.

\subsection{Engagement Estimation from Facial Expressions}
The rules for this score generation are derived from previous works~\cite{wang2010google,ko2019learning,joho2011looking} that correlate expressions with video types. Assume that $k$ defines the Genre ID ($k=0$ if the genre is \textit{comedy}, $k=1$ if the genre is \textit{tragedy}, $k=2$ if the genre is \textit{Anger}, $k=3$ if the genre is \textit{Horror}). Assume that an array $E[4]$ stores the total expression count for the $4$ expression categories\footnote{We consider \textit{Sadness} as \textit{Neutral}, as discussed earlier.}, as predicted during the content viewing. Let $R$ denote the number of times the expressions have changed during the content length of $l$ minutes. Let the function \texttt{indexOf} return the index of an element from the array $E[]$ ($-1$ if the element doesn't exist in the array), and $\land$ denote the logical AND operator. Then, the engagement indicator ($\mathcal{E}$) is estimated as follows.   
\begin{enumerate}
\item \underline{$\mathcal{E}=\text{True}$ if $k=\text{\texttt{indexOf}}(\max(E))$}: The strongest and most frequent expression matches with the content genre. For example, if the subject was mostly happy while watching a comedy video, it implies that the subject was engaged in the video.
\item \underline{$\mathcal{E}=\text{True}$ if $k \ne \text{\texttt{indexOf}}(\max(E)) \land \text{\texttt{indexOf}}(\max(E))=1 \land E[k] > \text{\texttt{avg}}(E[0], E[2], E[3])$}: This formula is based on the observation that even if a person is engaged, they might stay \textit{Neutral} for most of the time and occasionally show an expression that matches with the genre. It implies that even if the most frequent expression is neutral, the expression matching the content genre should be greater than the average of all expressions (other than neutral) found during the course. For the genre of tragedy, we check if the frequency of neutral (or Sadness) is greater than the average of all expressions predicted.
\item \underline{$\mathcal{E}=\text{True}$ if $k \ne \text{\texttt{indexOf}}(\max(E)) \land \text{\texttt{indexOf}}(\max(E))=1 \land R>0.3\times l)$}: This formula is based on the observation that if a person is engaged, and neutral for most of the time, they will show at least some changes in the expression during the content viewing, to ensure that they are not blankly staring at the screen. However, this condition fails if some auxiliary task in the background causes a difference in the expression. 
\item $\mathcal{E}=\text{False}$, if none of the above conditions hold true.
\item For the content of mixed genre, where there is no predetermined expression to compare the predictions with, facial expression alone cannot be used to detect the person's engagement. Hence, for mixed-type contents, this score is NULL.
\end{enumerate}

\subsubsection{Computing the \revise{Engagement} Score}
\revise{Engagement} score is taken as the percentage of the expression that matches with the genre of the content, concerning all expressions for the genre ``Tragedy'' or with respect to all non-neutral expressions for all other genres. The following formula generates the score.
\begin{equation}
A=\begin{cases}
\frac{E[k]}{\sum_0^nE[n]} \times 100 & \text{such that $n=0,2,3$ if $k \ne 1$}\\
\frac{E[k]}{\sum_0^nE[n]} \times 100 & \text{such that $n=0,1,2,3$ if $k=1$}\end{cases}
\end{equation}
For mixed genre, the \revise{engagement} score is displayed as the percentage of each found expression.



\subsection{Experimental Setup}
\revise{We recruited $12$ volunteers for this study. $10$ of them (P1-P10) were the same participants who volunteered in the previous experiment (Subsection \ref{coreMeth}). As discussed earlier, \ourmodel{} works best as a user-dependent model. Therefore, to test the system's performance for new subjects, we further considered two additional participants (P11, P12) -- one male being an IT Professional and one female teacher between $31$ and $52$ years of age, respectively.} 

\subsubsection{Content} 
For this study, we considered $15$ YouTube videos from the categories of \textit{Comedy}, \textit{Tragedy}, \textit{Anger}, \textit{Horror} and \textit{Mixed} for testing \ourmodelv. Each category contained three videos of different duration. The short videos had an average duration of $7$ minutes, medium ones had an average duration of $15$ minutes, and long videos had an average duration of $30$ minutes. We intentionally avoided the usage of any video longer than $45$ minutes to avoid the possibility of a major involuntary drop in the sustained attention level of the participants due to disinterest. To ensure that the videos were engaging, their ratings were considered for selecting them. Most of the videos had a YouTube rating of more than $25$k. The comedy videos consisted of clips from Mr. Bean and other popular movies; tragedy videos were tagged under ``Sad stories", \textit{Horror} videos ranged from short stories to long animated horror stories. Mixed-type videos were selected so that they could not be categorized as any of the categories distinctly. For example, a tutorial on ``Machine Learning", a documentary on the best historical places in the world, or a video explaining Drake's Equation. All these videos could give rise to amazement, amusement, surprise, confusion, or no emotion at all. The horror stories had an element of surprise, like ``Jump scares'' that were not only meant to invoke fear but also surprise. While these four categories had distinctive characteristics, it was difficult to choose videos under the category of ``Anger" due to the inadequacy of videos under such a tag and also due to the uncertainty of the emotion the videos invoke amongst the viewers. For example, a video displaying a major social issue might cause anger in one viewer during empathy in another. We carefully selected three videos on social injustice like animal cruelty, bullying, and elder abuse that are likely to cause anger in the viewer. \revise{ Since YouTube videos have been found to cause expressions like anger~\cite{lee2012exploring}, the selection of videos for this category was also based on the selection of representative emotional terms \cite{chen2017emotion} like ``angry,'' ``force,'' etc. from the video's title, and comments~\cite{savigny2017emotion}\cite{sarakit2015classifying}.}


\subsubsection{Methodology} \label{meth}
The total experiment was conducted in two non-consecutive sessions -- an initial training session of $15$ minutes and a session for testing \ourmodel streaming app. In the first session, the entire experiment was explained to the participants. \revise{The participants could choose to watch a video using \ourmodel{} streaming app to understand its basic functionality. }  To eliminate forced attention, we assured the participants that there would be no penalty for lower scores of \revise{engagement} or disengagement. After the training session, the participants were asked to use the \ourmodel{} streaming app for the second session. This session $s_2$ was divided into $5$ sub-sessions -- $s_2^1$... $s_2^5$, where $s_2^1$ was dedicated to viewing the $3$ videos from the genre Comedy, $s_2^2$ was dedicated to tragedy and so on. Within each sub-session $s_2^n$, the participant had to watch $3$ videos of different duration and were interviewed after each video. The participants were free to take breaks (at least for $15$ mins) within or between these sub-sessions to remove eye fatigue. Each sub-session $s_2^n$ was planned to have a duration of $1.5$ hours as the total viewing time for short, medium, and long videos were $\approx60$ minutes, and the total interview time was $30$ minutes. However, considering the breaks within the sub-sessions, the maximum time was increased to $2$-$3$ hours. Hence, for each participant to complete viewing all the videos in \ourmodel{} streaming app from each of the $5$ categories, a total of $10$-$15$ hours were estimated. Each participant completed these experiments over multiple consecutive days based on their interests. Also, they were free to choose which video they wanted to see at a point in time from the list of available videos. We delegated these choices to the participants to ensure that they could watch the videos freely and that watching consecutive videos does not put much cognitive load on them, which might affect their \revise{engagement} level. 

There was no restriction on the sitting position of the participants; however, the placement of the device was kept optimal based on the viewing preference of individual participants. The participants were requested to minimize significant body movements during the sessions, and the ambient noise was minimal to avoid disturbance. In addition, we added secondary tasks like showing funny, sad, and scary images (that did not match with the instantaneous video genre) on a screen behind the device for Participant $7$ during the experimental sessions. This was to test if \ourmodel{} streaming app could also capture low \revise{engagement} levels caused by secondary tasks. \revise{For each video genre, the predictions of \ourmodel{} were generated for a total of $\approx$ 930 data points while each participant watched the short ($\approx$ 126 data points), medium  ($\approx$ 270 data points) and long  ($\approx$ 540 data points) videos.}

\subsubsection{Interview Mechanism}
After each experimental session, the interview consisted of two question-answer sessions.\\

\noindent(1) \textit{Questions from the video content}: These questions were asked based on the video viewed before the interview and were selected in such a way that captured the overall \revise{engagement} of the participant. For example, for a $30$ minutes long video, the first question was based on the first $3$ minutes of the video, the second question was from $3$rd to $6$th minute of the video, and so on. For short videos, medium, and long videos, there were two, five, and ten questions, respectively. The questions did not come with options to avoid the possibility of guessing. Based on whether the participant answered the question correctly, we marked it as $0$ or $1$. The final \textbf{ground truth \revise{engagement} score} was estimated by taking the percentage of the correct answers for each video under each category. \revise{A total of $85$ questions were asked to each participant for all the videos and genres combined.} \\ 

\noindent(2) \textit{Self-assessment}: Self-perception of engagement refers to whether the user considers themselves engaged or disengaged. Self-perception is essential in the case of content rating. For example, if they feel that the content is engaging, their ratings of the content will be high. Suppose the estimation generated by \ourmodel{} can relate to this personal or self-assessment of engagement. In that case, it can promote automated feedback by eliminating the requirement of manual rating or feedback, as manual feedback can be biased or influenced by different compelling factors\footnote{\url{https://factorialhr.com/blog/bias-in-performance-reviews/\#types-of-bias-in-performance-reviews} (Accessed: \today)}. Whether the user perceives themselves as engaged depends on two underlying factors -- Whether the viewer pays attention to the content (F1) and whether the viewer found the content interesting (F2); in some cases, F2=F1. For example, if the user is bored with the content, their sustained attention will drop. Conversely, exciting content will promote engagement. In other cases, F1 and F2 can be unrelated. For example, the user may find the content boring but still choose to pay attention to it. Thus, it is essential to analyze each of the two factors independently to understand the overall engagement of the user. 

Based on this understanding, after each video, the participants were asked if they paid attention to the video shown in the application or were focused on something else. They were also asked if they found the video interesting. These two questions were yes(1)/no(0) type. The \textbf{self-engagement \revise{indicator}} for each video was calculated by taking the logical AND between these two answers.



\subsection{Hypotheses}
This study hypothesizes that \ourmodel-generated scores will highly correlate with the ground truth \revise{engagement} score and the self-engagement \revise{indicator}. From theoretical evidence, we also hypothesize that most facial expressions will be Neutral, as, in real-world scenarios, content-invoked expressions are sparse and aperiodic. \revise{Therefore, changes in the expressions would be natural depending on the video content, and the models should still be able to correlate with the manual (ground truth \revise{engagement score} and self-engagement \revise{indicator}) scores. If we observe a high accuracy in this prediction, \ourmodel{} could likely identify the natural expressions correctly during the session.}

\subsection{Results}
In this subsection, we discuss the performance of \ourmodelv{} through the analysis and comparison of \revise{engagement} scores and engagement \revise{indicators}.

\begin{figure}[!ht]
   \includegraphics[width=.8\textwidth]{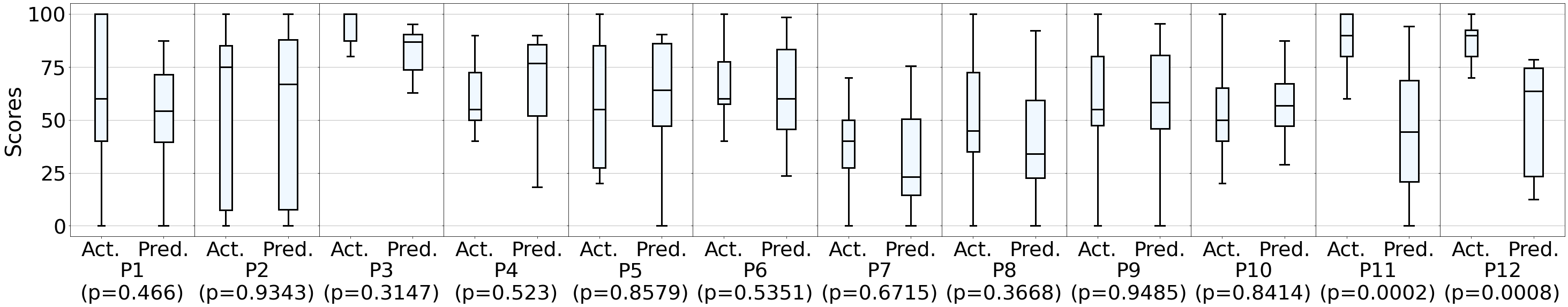}
  \caption{Distribution of participant-wise actual and predicted \revise{engagement} scores in \ourmodelv. The null hypothesis is that the two score distributions are similar. The graph shows that, for P1-P10, the null hypothesis is accepted (p>0.05). For P11-P12, the alternate hypothesis that the scores are different, is accepted.  }
  \Description{The figure shows a box and whiskers graph, comparing the actual and predicted scores of the 12 participants.}
  \label{vidpwise}
\end{figure}

\subsubsection{Analysis of \revise{Engagement Score}}
Figure~\ref{vidpwise} depicts the distribution of the ground truth \revise{engagement} scores generated from the interview with that of the scores generated by \ourmodel{} streaming app. The graph compares the distribution of these two scores for each participant by considering all the videos from the $4$ video genres -- comedy, tragedy, anger, and horror. We infer the following observations from these results. 
\begin{enumerate}
    \item The ground truth and the predicted scores are significantly correlated (with the reported p-values of a statistical T-test between the two distributions) for most of the participants, thus proving a part of our hypothesis.
    \item For P7, it can be seen that both the ground truth and the predicted scores are low. This is because P7 paid attention to the secondary task, which caused a variety of facial expressions (based on the images), thus leading to a lower predicted score. Similarly, for P3, both these scores are high as the participant was fully \revise{engaged} to the videos and could answer most of the questions correctly. This proves the capability of \ourmodel{} streaming app to distinguish \revise{engaged} participants from the less-\revise{engaged} ones, thus indicating the success of \ourmodel{} for correctly inferring the natural facial expressions.
    \item For P11 and P12, the difference in the ground truth and predicted scores were caused by the misclassification of expressions by \ourmodel. This is because the system is user-dependent and needs to be calibrated and trained with a few data samples from the users before being able to perform with significant accuracy. However, even though the error level for unseen participants was high, we could observe a correlation between the score level. For higher ground truth scores, the overall distribution of the predicted scores tends to be higher.
\end{enumerate}
\revise{This observation leads to the inference that affective facial expressions are correlated to the attention level of a person~\cite{10.1145/3555656}. The correlation can be estimated by analyzing its similarity with the video genre.}

\begin{figure}[!ht]
   \includegraphics[width=.35\textwidth]{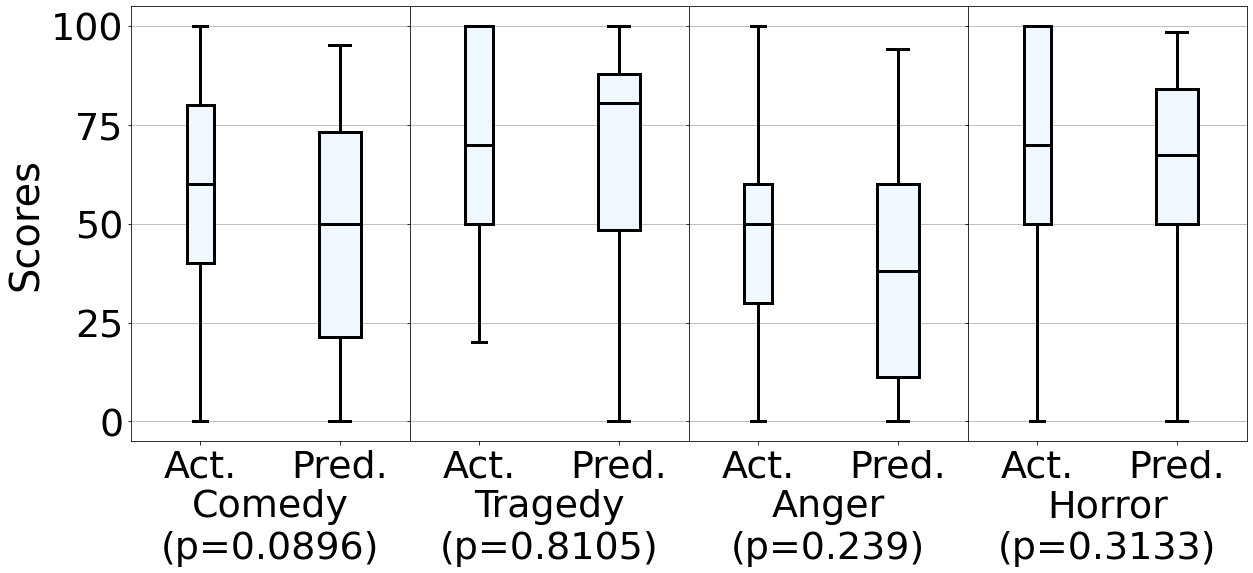}
  \caption{Distribution of genre-wise actual and predicted \revise{engagement} scores; the null hypothesis is that the two score distributions are similar. The graph shows that, for all genres, the null hypothesis is accepted (p>0.05).}
  \Description{The figure shows a box and whiskers plot comparing the actual and predicted scores for each video genre. The null hypothesis is that the two score distributions are similar. The graph shows that, for all genres, the null hypothesis is accepted (p>0.05). }
  \label{vidgwise}
\end{figure}

Next, Figure~\ref{vidgwise} shows the genre-wise distribution of the ground truth \revise{engagement} score and the predicted scores. We observe a strong correlation between the two scores indicating that \ourmodel{} streaming app can perform well for various video types. However, the overall score under the anger category is lower than that of others. This is because anger videos failed to invoke the expression of anger in most of the participants. Moreover, the participants were found to be less attentive to such videos in this study. On the contrary, the expressions were mainly neutral for tragedy, but the attention levels were high. This aligns with the interchangeability of sadness and neutral expressions in \ourmodel. However, it was noted that some of the tragedy videos received ``empathetic smiles'' from the users. Such instances were mis-classified as ``Happy'' thus leading to some level of disagreement between the video genre and found expressions. \revise{By comparing these scores based on different thresholds, it was found that the average F1-score for optimal thresholds in each video category is $.84$.}

\subsubsection{Analysis of Engagement for Mixed Genre}
\begin{figure}[!t]%
    \centering
    \subfloat[\centering Short videos]{{\includegraphics[width=5cm]{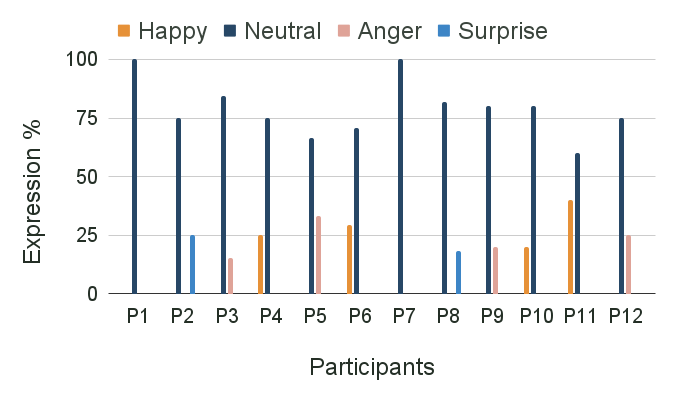} }}%
    \subfloat[\centering  Medium videos]{{\includegraphics[width=5cm]{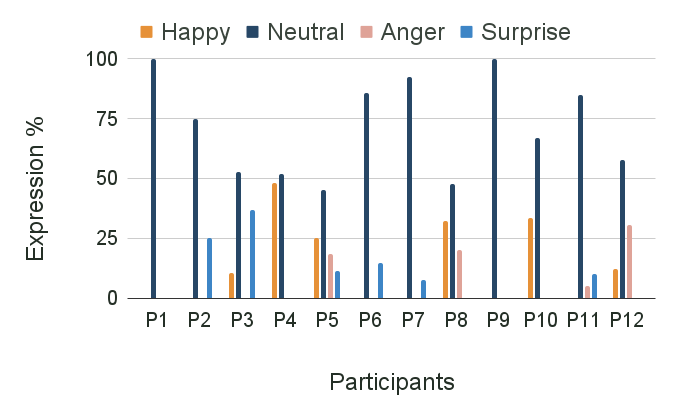} }}%
    \subfloat[\centering  Long videos]{{\includegraphics[width=5cm]{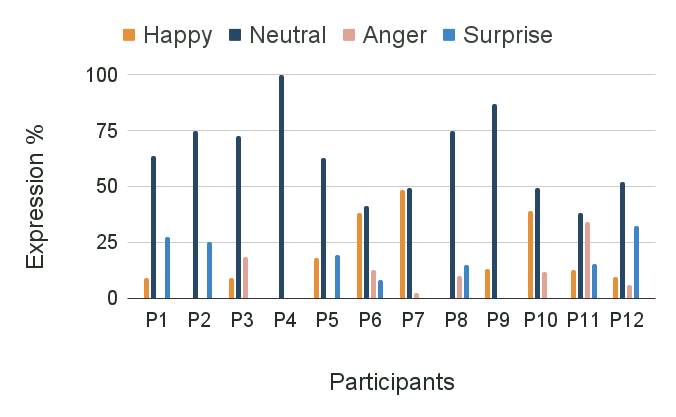} }}%
    \caption{Distribution of facial expressions for different mixed type videos for individual participants}%
    \Description{The figure shows three bar graphs, each for short, medium and long mixed type videos. The graphs show the number of expressions found for each participant P1-P12 for these videos. It can be seen that for long videos, the number of expressions has increased. }
    \label{mixvid}%
\end{figure}

As discussed earlier, for mixed videos, since there are no pre-determined regular expressions, \revise{engagement} score based on one single expression is rather unfair. Hence, we compare the percentages of all the found expressions for these videos for each participant for the three different videos of lengths short, medium, and long. Figure~\ref{mixvid} graphically shows these distributions for different video duration. The following inferences can be drawn from the figure. 
\begin{enumerate}
    \item As hypothesized, for almost all the participants and video duration, the most prevalent expression is found to be neutral.
    \item A clear correlation between the video length and the number of expressions can be established. It can be seen that for medium or longer videos, the participants showed a greater number of expressions than for short videos. This indicates that for mixed-type videos of longer duration, \revise{engagement} can be estimated by mapping the user's expression with other video viewers.
\end{enumerate}

\subsubsection{Analysis of Engagement \revise{Indicator}}
\begin{figure}[!t]
  \begin{minipage}[b]{0.5\textwidth}
   \includegraphics[width=\textwidth]{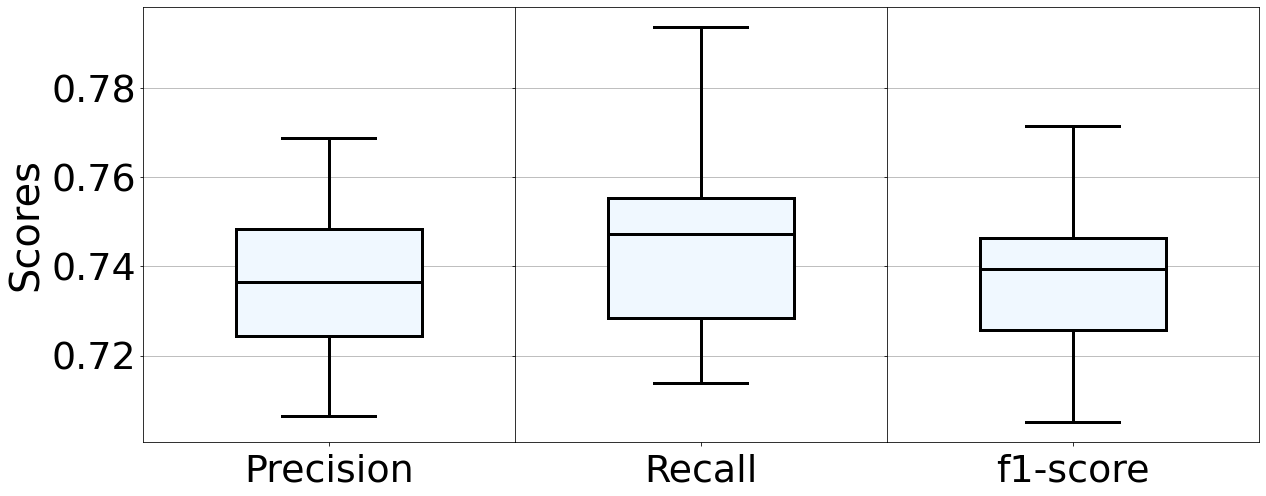}
  \caption{Comparison of Overall Precision, Recall and f1-score for self reported engagement \revise{indicator} vs predicted engagement \revise{indicator} for \ourmodelv}
  \Description{The box and whiskers plot shows that the precision, recall and f1-score for self reported and predicted engagement vary between 0.7 to 0.8. }
  \label{videng}
 \end{minipage}
  \hfill
  \begin{minipage}[b]{0.45\textwidth}
   \includegraphics[width=\textwidth]{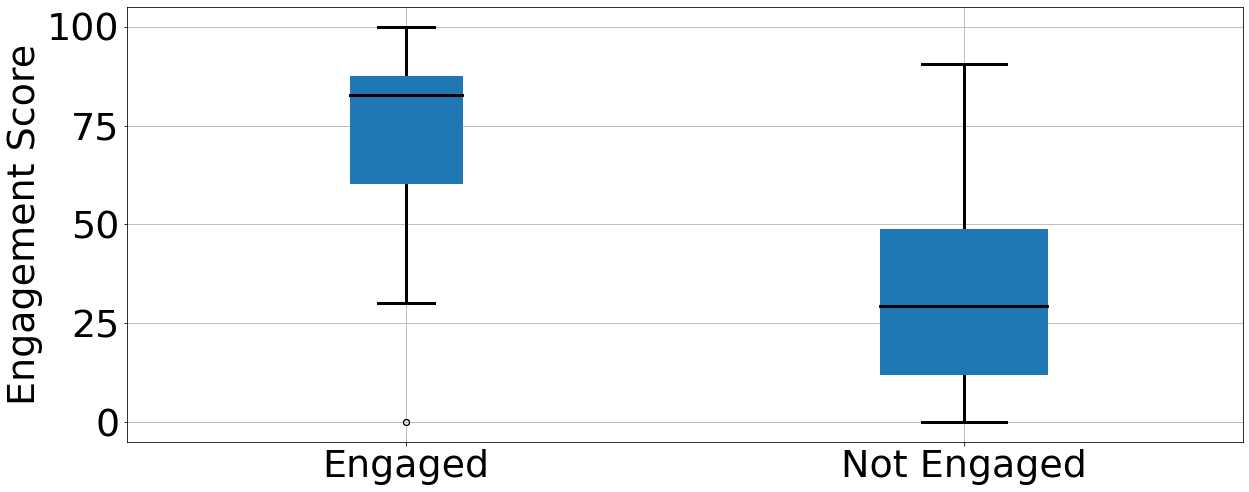}
  \caption{Correlation between \revise{engagement score} and engagement \revise{indicator}, as predicted by  \ourmodelv. The null hypothesis that the distribution of these scores are similar is rejected by the ttest as the p-value $<<$ 0.05.  }
  \Description{This box and whisker plot shows that for videos where the participants were indicated to be engaged, the \revise{engagement} scores varied between 60 to 95 and those indicated as not engaged had \revise{engagement} scores between $\approx$20 to 50.}
  \label{vidaten}
    \end{minipage}
\end{figure}

We now explore \ourmodel{} streaming app's performance in terms of engagement indicator. In Figure~\ref{videng}, we compare the engagement labels as predicted by the \ourmodel{} streaming app with the self-assessment score. The graph shows a significant correlation between the predicted score and the self-assessment score. It aligns with the assumption that engagement is the underlying attention and interest of the user to the video content and that it can be quantified by considering the strength of expressions and their rate of change. Even though engagement \revise{indicator} and \revise{engagement} scores are estimated as separate variables, Figure~\ref{vidaten} shows that if a user is marked as engaged by \ourmodel{} streaming app, they are likely to have higher \revise{engagement} scores. In contrast, disengaged users will usually have significantly lower \revise{engagement} scores.

\section{Large-scale Usability Study with \ourmodel{} Streaming App}
\label{usability}
\begin{figure}[!ht]
  \begin{minipage}[b]{0.45\textwidth}
   \includegraphics[width=0.8\textwidth]{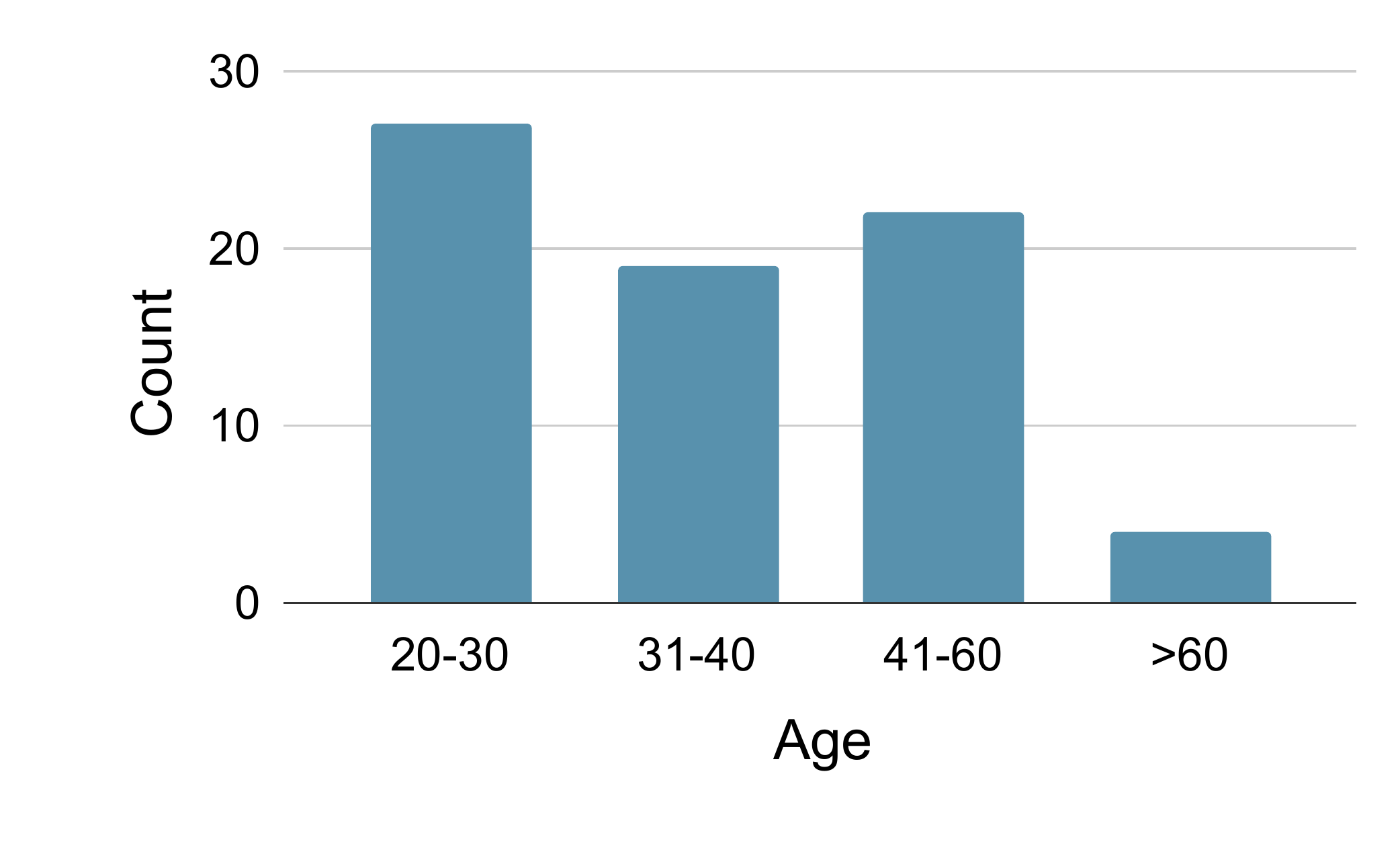}
  \caption{Distribution of participants based on age groups}
  \Description{The bar plot shows that 45\% of the participants belonged to the age group of 20-30 years, 30\% to 31-40, 19\% to 41-60 and the rest were above 60.  }
  \label{page}
\end{minipage}
  \begin{minipage}[b]{0.45\textwidth}
   \includegraphics[width=0.8\textwidth]{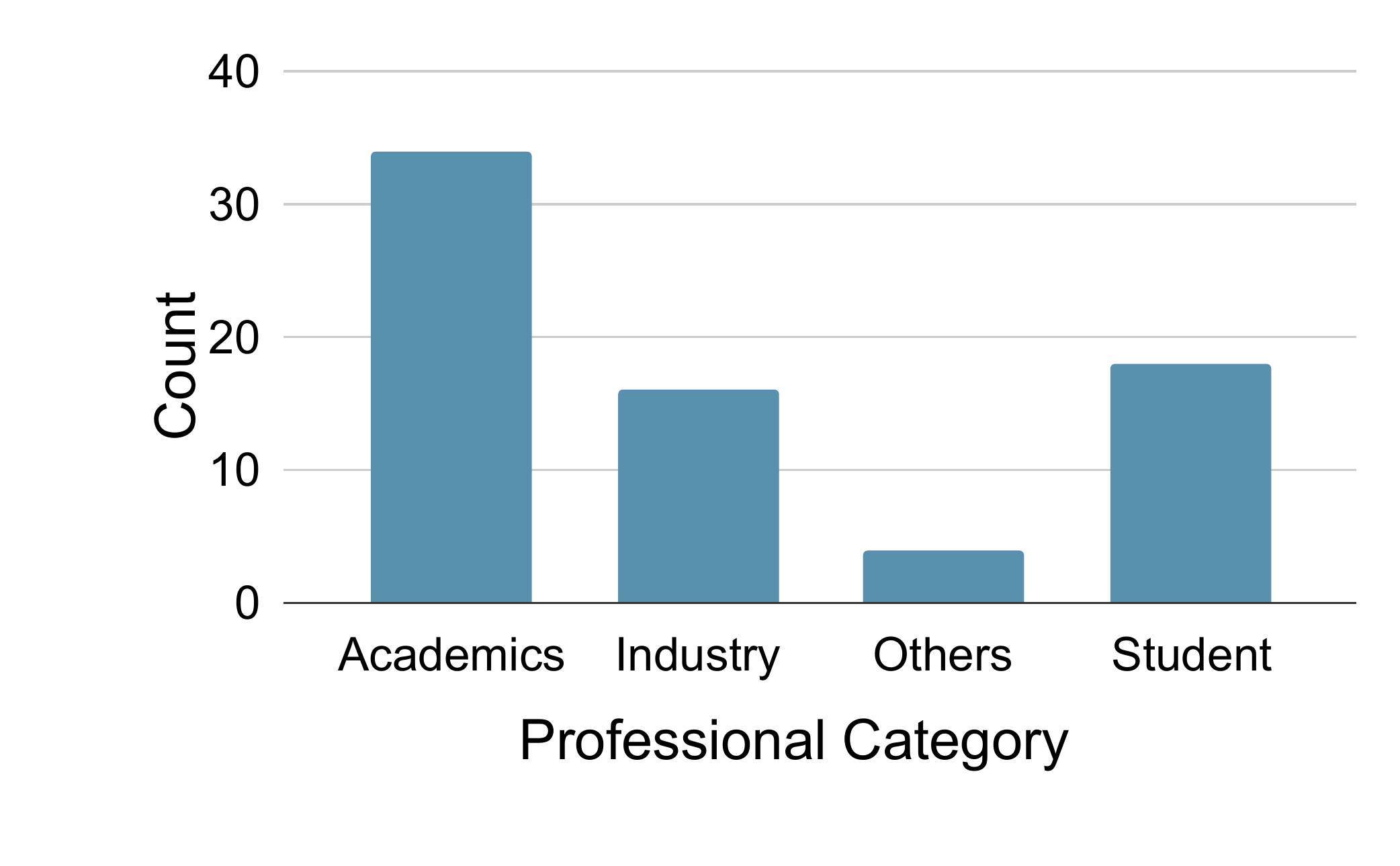}
  \caption{Distribution of participants based on profession}
  \Description{The bar plot shows that 34 participants belong to Academics, 16 from Industry, 4 from Others and the rest are Students. }
  \label{pprof}
\end{minipage}
\end{figure}

The objective of this study is to analyze how well the participants in-the-wild rate \ourmodel{} streaming app in terms of its usability in the practice. In this study, we have considered $72$ subjects from different countries, professions and age groups (between $18$-$71$). The participants were from different countries like United States ($10$), India ($27$), South Korea ($4$), Germany ($3$), United Kingdom ($2$), Australia ($5$), Bangladesh ($3$), Japan ($3$), Austria ($1$), Brazil ($2$), Canada ($3$), Croatia ($1$), Israel ($1$), Hong Kong ($3$), and China ($4$). $47$ of the participants were male, and the rest were female. While Figure~\ref{page} shows the distribution of age-groups and the corresponding count of participants, Figure~\ref{pprof} shows the professional categories to which the participants belonged. The category ``\textit{Academics}'' comprises of professions like Professors, Research Scientist, Research Scholars, Post Doctoral Fellows and Educators. We have grouped professions like Service, Software Engineers, Software Developers and IT under ``\textit{Industry}''. Banking Services and Government employees are categorized  under ``\textit{Others}''. It can be noted that these participants are distinct from those who participated in the previous studies. \revise{Notably, we did not have the ground-truth information for these participants, so we only analyzed the usability ratings they submitted after experiencing the app.}

\subsection{Methodology}
In this study, we released the APK of \ourmodel{} streaming app to the participants, along with a video demonstrating the applications. The participants were requested to install the applications, use them thoroughly and provide their feedback through a Questionnaire based on the System Usability Scale (SUS)~\cite{brooke1996sus}. The questionnaire is a series of positive and negative questions, presented in an alternate manner. In this survey, the odd numbered statements are the positive ones, and even numbered ones are the negative ones and the participants can provide a score between $1$--$5$ to each statement based on their level of disagreement ($1$) or agreement ($5$) to the statement. The statements are as follows.
\begin{enumerate}
    \item[Q1.] I think that I would like to use this system frequently.
    \item[Q2.] I found the system unnecessarily complex.
    \item[Q3.] I thought the system was easy to use.
    \item[Q4.] I think that I would need the support of a technical person to be able to use this system.
    \item[Q5.] I found the various functions in this system were well integrated.
    \item[Q6.] I thought there was too much inconsistency in this system.
    \item[Q7.] I would imagine that most people would learn to use this system very quickly.
    \item[Q8.] I found the system very cumbersome to use.
    \item[Q9.] I felt very confident using the system.
    \item[Q10.] I needed to learn a lot of things before I could get going with this system.
\end{enumerate}
The final SUS score is calculated as :\\ 
\texttt{((QA1-1)+(5-QA2)+(QA3-1)+(5-QA4)+(QA5-1)+(5-QA6)+(QA7-1)+(5-QA8)+(QA9-1)+(5-QA10))*2.5},\\ 
where \texttt{QAn} is the score to statement \texttt{Qn}, provided by a participant. 
 
\subsection{Result}
\begin{figure}[!ht]
  \begin{minipage}[b]{0.55\textwidth}
   \includegraphics[width=0.75\textwidth]{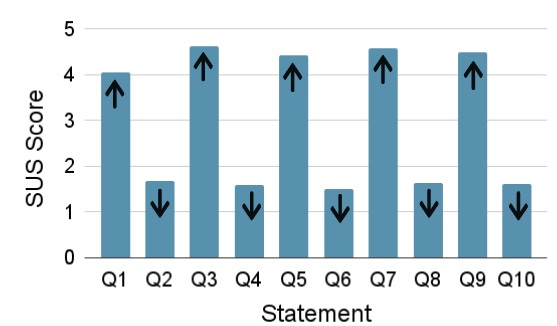}
  \caption{Distribution of SUS scores based on individual statements}
  \Description{A bar plot shows that the odd statements of SUS survey have received scores more than 4 while the even statements have received scores less than 2. }
  \label{susall}
\end{minipage}
  \begin{minipage}[b]{0.35\textwidth}
   \includegraphics[width=\textwidth]{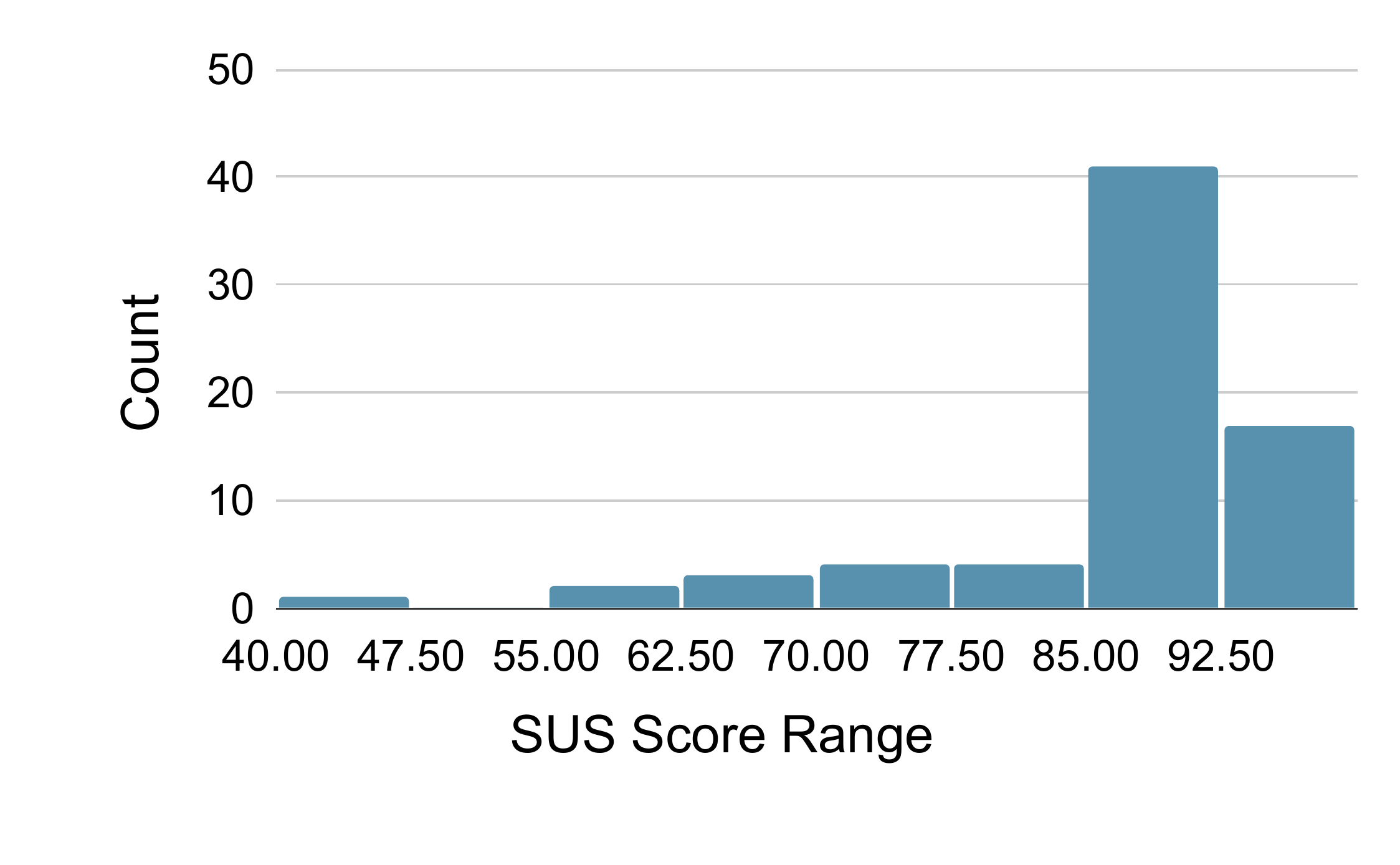}
  \caption{Histogram of SUS Scores}
  \Description{A histogram that shows that about 30 of the responses indicated a SUS score of more than 80.}
  \label{sushis}
\end{minipage}
\end{figure}

From this experiment, we received an average SUS score of $85.34$ which establishes the usability of a system like \ourmodel. To further test our hypothesis, we plot Figure~\ref{susall} that shows the questionnaire's statement-wise average score (on the scale of $1$--$5$), as provided by the participants.
Figure~\ref{sushis} shows that the majority ($61$) of the responses indicated an SUS score $80$ or more. We also found that age of the participants had no effect on the usability of the system. The age-group-wise aggregation of the scores varied between $82.1$ to $90$, showing the application was equally usable by people from all  age groups. Similarly, for both male and female users, the average SUS score was similar ($84.5$ for males and $86.02$ for females). Based on professions, the application was rated highest by academicians ($87.5$), followed by users from industry ($86.09$), students ($81.1$) and other services ($80.6$). The application was also found to be widely accepted by people from various cultural backgrounds. The average score based on the demographic information of the participants showed that for different countries, the SUS score ranged from $80.1$ to $91.25$.

\section{Discussion}
\ourmodel{} demonstrates how a single microphone and a single speaker present in a commercial smartphone can address the problems of camera-based facial expression detection. However,  our model also suffers from a few limitations that make it suitable as a performance enhancer of camera-based techniques. However, in scenarios with a limited number of expression variations, \ourmodel{} can even replace camera-based techniques. This section discusses some of the limitations and future scopes of \ourmodel.
\subsection{Near-ultrasonic nature of the signals}
The supported frequency range is up to $20$ kHz in a commodity smartphone; the near ultrasound signals used in this model are slightly audible. This can cause mild discomfort in some users. As reported by some of the participants under feedback, the chirps, though mildly audible, created minor distractions. However, this problem can be mitigated by periodic system usage or by further narrowing the frequency range. In the future, we will aim to extend the system to the iOS platform as some iPhones can support frequencies above $20$ kHz.
\subsection{Effects of obstruction, movement and device orientation}
Though the current experiments allow natural body movements, they lack large-scale movements. The performance of \ourmodel{} in the presence of significant activity is subject to further testing. However, such noises can easily be eliminated by sudden and abnormal changes in I-Q values, measured against a manual threshold. Moreover, since we use smartphones, it has been assumed that the face will be the nearest object to the device, and the reflected signals from the facial regions will be the strongest. However, the performance of \ourmodel{} might degrade if an additional obstruction (like a mask or spoon) is inserted between the device and the face. \revise{ The current prototype of \ourmodelv{} is developed to work in portrait (vertical orientation). Even though some users might prefer to hold the device in landscape mode, it would negatively impact the accuracy of \ourmodelv{}. This is due to the fact that in this orientation, when a person hold the smartphone the microphone (as well as the speaker) will be mostly covered by the user's palm (based on the observation in section \ref{handp}).    }
\subsection{Validity of engagement scores}
In this work, due to the lack of any standard metric, we have compared the predicted engagement score with that generated manually through interviews and questions based on the viewed content. However, such manual scores can sometimes be misleading. Since the interviews take place at the end of each video/reading session, for longer content, the users could forget the answer to a question based on the first part of the video or story. This might lead to a lower manual engagement score and a high predicted score based on continuous monitoring of expressions. However, we assume that the contents are of optimal duration, such that if a participant pays engagement, the information will be retained in their short-term memory. Moreover, misclassified expressions can be generated by \ourmodel, which might create lower scores. In the future, we aim to extend the data for training the model to achieve higher accuracy.


\section{Conclusion}
In this paper, we present \ourmodel, a lightweight near-ultrasound signal-based facial expression detection system that works in real-time on a commodity smartphone. The system not only eliminates the requirement of any additional hardware or modification of the inbuilt components of any inexpensive smartphone with minimal processing capacity but also overcomes the challenges associated with camera-based techniques -- such as occlusion or privacy impairment. Through rigorous lab-scaled and unconstrained testing, \ourmodel{} depicts a significant performance in classifying various facial expressions and proves the application of acoustic sensing in this domain. It also reveals the capabilities of commercial smartphones in facilitating such acoustic applications, thus proving its feasibility and scope of global acceptance.  
\bibliographystyle{ACM-Reference-Format}
\bibliography{main}
\end{document}